\documentclass[12pt]{article}
\usepackage{amsmath}
\usepackage{graphicx}
\usepackage{enumerate}
\usepackage{natbib}
\usepackage{amssymb}
\usepackage{amsthm}
\usepackage{xcolor}
\usepackage{CJK}
\usepackage{blkarray}
\usepackage{mathrsfs}
\usepackage{booktabs}
\usepackage{epstopdf}
\usepackage{diagbox}
\usepackage{amsfonts}
\usepackage{multicol}
\usepackage{multirow}
\usepackage{algorithm}
\usepackage{subcaption}
\usepackage{algorithmic}
\usepackage{rotating}
\usepackage{setspace}
\usepackage{graphicx}
\usepackage{verbatim} 
\usepackage[bookmarks=false,
colorlinks,
linkcolor=blue,
anchorcolor=blue,
citecolor=blue,
filecolor=blue,
urlcolor=black
]{hyperref}

\newtheorem{thm}{Theorem}

\newtheorem{df}{Definition}
\newcommand{\dT}{\mathsf{T}}

\def\squarebox#1{\hbox to #1{\hfill\vbox to #1{\vfill}}}
\def\boxit#1{\vbox{\hrule\hbox{\vrule\kern6pt
			\vbox{\kern6pt#1\kern6pt}\kern6pt\vrule}\hrule}}

\allowdisplaybreaks

\date{}
\def\spacingset#1{\renewcommand{\baselinestretch}%
	{#1}\small\normalsize} \spacingset{1}

\begin{document}

\title{\bf Delaunay Weighted Two-sample Test for High-dimensional Data by Incorporating Geometric Information}
\author{Jiaqi Gu\footnote{jiaqigu@usf.edu; ORCID: 0000-0002-8773-9558}\\
Department of Mathematics and Statistics, University of South Florida \\
and \\
Ruoxu Tan\footnote{Corresponding author, ruoxut@tongji.edu.cn; ORCID: 0000-0002-2046-4377} \\
School of Mathematical Sciences and \\
School of Economics and Management, Tongji University\\ 
and \\
Guosheng Yin\footnote{gyin@hku.hk; ORCID: 0000-0003-3276-1392} \\
Department of Statistics and Actuarial Science, University of Hong Kong\\ }

\maketitle

\begin{abstract}
	Two-sample hypothesis testing is a fundamental problem with various applications, which faces new challenges in the high-dimensional context. To mitigate the issue of the curse of dimensionality, high-dimensional data are typically assumed to lie on a low-dimensional manifold. To incorporate geometric information in the data, we propose to apply the Delaunay triangulation and develop the Delaunay weight to measure the geometric proximity among data points. In contrast to existing similarity measures that only utilize pairwise distances, the Delaunay weight can take both the distance and direction information into account. A detailed computation procedure is developed to learn the unknown manifold and approximate the Delaunay weight. We further propose a novel nonparametric test statistic using the Delaunay weight matrix. Asymptotic normality under the null and consistency under the alternative of the test statistic are developed. Applied to simulated data, the new test shows robustness to the learning of the unknown manifold and exhibits substantial power gain if the distributions differ in the principal directions of covariance matrices. The proposed test also detects significant differences on a real dataset of mice protein expression levels.
\end{abstract} 

\spacingset{1.9}

\section{Introduction}\label{Introduction}

The two-sample test on whether the two underlying distributions are the same or not is a fundamental problem in statistics and machine learning with broad applications in change point detection \citep{Cao2018,Chen2019}, goodness-of-fit evaluation \citep{Chwialkowski2016,AriasCastro2018} and experimental design \citep{Morgan2012}. In the era of big data, there are emerging challenges that the power of classical two-sample tests (e.g., Hotelling's $T$-squared test, Wald test or likelihood ratio test) decreases or even vanishes as the data dimension increases. Under the high-dimensional setting, several methods have been developed to test a simpler hypothesis on whether two population means are the same. For example, under the sparse alternative that mean vectors of two samples only differ in a small number of dimensions, \citet{Cai2013a} developed a maximum-type test statistic. To enhance the power and robustness to extreme values in detecting sparse mean differences, a parametric bootstrap approach with feature screening was proposed by \citet{Chang2017}, while \citet{Wang2019} incorporated the trimmed mean approach to developing a robust test. \citet{Liu2022} developed a projection test by assuming the sparsity of the optimal projection direction. \citet{Yang2024}
introduced a novel test whose estimation error does not affect asymptotic power while achieving superior performance. 
However, the aforementioned methods are not consistent against general alternatives, e.g., the two populations possess the same mean but differ in the covariance structures.

Nonparametric two-sample testing for general differences between distributions is a historical research topic with rich literature. For univariate data, the Kolmogorov--Smirnov test \citep{Kolmogorov1933,Smirnov1948}, the Wilcoxon--Mann--Whitney test \citep{Wilcoxon1945,Mann1947} and the Wald--Wolfowitz runs test \citep{Wald1940} are early developments on this topic. Although such methods may be directly applicable to multivariate settings \citep{Bickel1969}, they would perform poorly unless the total sample size grows at an 
exponential rate with  
the dimension. To  
circumvent such difficulty, \citet{Friedman1979} developed a two-sample test based on the minimum spanning tree (MST) that connects all data points with the minimal total length of edges. Given an MST of the pooled sample, the test statistic is constructed as the number of edges connecting data points from different samples, which is shown to be consistent against general alternatives. More recently, many nonparametric two-sample tests have been proposed for multivariate/high-dimensional data, including:
(i) distance/energy-based tests based on pairwise distances of the pooled sample \citep{Szekely2004,Jureckova2012,Marozzi2015}; (ii) graph-based tests, e.g., the $k$-nearest neighbor ($k$-NN) graph \citep{Schilling1986,Henze1988,Hall2002} and the $k$-MST graph \citep{Chen2017,Chen2018};
(iii) kernel tests via a decaying kernel of pairwise Euclidean distances of the pooled sample \citep{Gretton2009,Gretton2012,Cheng2024,Yan2023}; and
(iv) regression tests by connecting the binary classification and two-sample testing \citep{Kim2019,Hediger2022}. 

To mitigate the curse of dimensionality, high-dimensional data are often assumed to  
lie on an unknown low-dimensional manifold, 
which is reasonable in many applications, such as bioinformatics \citep{Moon2018}, image data analysis \citep{Pless2009} and natural language processing \citep{Zhao2021}.  
Performances of the 
nonparametric tests under the low-dimensional manifold setting have been studied in the literature. For example, it is shown that the kernel test \citep{Cheng2024} and the local regression test \citep{Kim2019} are adaptive to the intrinsic dimension of the manifold by properly choosing the tuning parameter. \citet{AriasCastro2018} also showed that the multivariate chi-squared test is adaptive to the intrinsic dimension of the manifold, while their work was mainly focused on theoretical development rather than practical implementation. 
However, none of these tests directly take the manifold structure into account, while our work aims to fill this gap.

In particular, we develop a new nonparametric two-sample test, named the Delaunay weighted test, for high-dimensional data under the low-dimensional manifold assumption. Following the $k$-NN test by \citet{Schilling1986}, we  propose the test statistic as the average of all within-group proximities of the pooled sample. The proximity is measured by our newly developed Delaunay weight instead of the $k$-NN. By incorporating the Delaunay triangulation on a Riemannian manifold \citep{Leibon2000}, the Delaunay weight measures the geometric proximity among data points by taking both the geodesic distance and the relative direction into account. In practice, the Delaunay weight cannot be computed exactly because the manifold is usually unknown. As a remedy, we compute the Delaunay weight matrix of low-dimensional manifold representations for approximate inference. The manifold representations are obtained following the idea of Isomap~\citep{Tenenbaum2000}. We develop the stereographic projected \textsf{DELAUNAYSPARSE} algorithm to find the Delaunay simplices for computing the Delaunay weight. The $p$-value of our test is obtained by a permutation procedure, and our Delaunay weighted test is applicable as
long as the sample size is larger than the (estimated) intrinsic dimension of the underlying manifold.  
Compared with the kernel test and the local regression test based on local distances that are adaptive to the intrinsic dimension of the manifold, ours further takes the relative direction among data points into account, and thus it is more efficient in detecting the direction difference of two distributions. Numerical experiments on synthetic data reveal that our Delaunay weighted test outperforms existing approaches when the two distributions differ in the principal directions of covariance matrices, although the power of our method is not the highest under location differences between distributions.

The rest of this article is organized as follows.  Section \ref{Preliminaries} motivates and formulates the two-sample testing problem for high-dimensional data lying on an unknown Riemannian manifold and introduces preliminary geometric concepts. The Delaunay weighted test is proposed in Section \ref{Method}, including the definition and advantages of Delaunay weight, the test statistic, computational strategies to approximate the Delaunay weight, and the 
permutation procedure for inference.  
Section \ref{Theory} investigates the theoretical properties of the proposed test. In Section \ref{experiments}, we conduct various numerical experiments including a real data example to demonstrate the practical performance of our method, such as the statistical power, robustness to the manifold learning procedures, and computational cost. Section \ref{Discussions} concludes with discussions. The
supplementary material contains details of our computational strategies, proofs of theorems and additional experimental results.

\section{Manifold Hypothesis and Delaunay Triangulation}\label{Preliminaries}

\subsection{Manifold Hypothesis}\label{Motivation}

Consider two samples $\mathbb{X}=\{\textbf{x}_1,\ldots,\textbf{x}_{n_1}\}\subset\mathcal{R}^D$ and $\mathbb{Y}=\{\textbf{y}_1,\ldots,\textbf{y}_{n_0}\}\subset\mathcal{R}^D$ of independent and identically distributed (i.i.d.) data from two distributions $F$ and $G$,
respectively. Our goal is to test $H_0:F=G$ versus $H_1:F\neq G$. Let $\mathbb{Z}$ be the union of two samples, $\mathbb{Z}=\mathbb{X}\cup\mathbb{Y}=\{\textbf{z}_1,\ldots,\textbf{z}_n\}$, and let $\boldsymbol{\delta}=\{\delta_1,\ldots,\delta_n\}$ denote the group indicator of the pooled sample, where $n={n_1}+{n_0}$,  $\delta_i=1$ if $\textbf{z}_i\in\mathbb{X}$ and $\delta_i=0$ if $\textbf{z}_i\in\mathbb{Y}$, for $i=1,\ldots,n$.  
    
We assume that both $F$ and $G$ are supported on a $d$-dimensional geodesically convex Riemannian manifold $\mathcal{M}$ embedded in the ambient space $\mathcal{R}^D$ with $d \ll D$. This assumption, also known as the manifold hypothesis, is reasonable in many real data examples and prevailing in high-dimensional data analysis \citep{Pless2009,Meilua2024}. For example, each image often contain several thousand pixels, while \citet{Gong2019} showed that the intrinsic dimension of a face image dataset is around 10 to 20. In addition, the manifold hypothesis has been invoked in analyses of bioinformatics \citep{Moon2018}, natural language processing \citep{Zhao2021}, etc.

The manifold hypothesis is also investigated in the literature of two-sample testing \citep{AriasCastro2018,Kim2019,Cheng2024}.
To illustrate how the manifold hypothesis is helpful in two-sample testing applications, we consider an example of biological data analysis. In biological experiments, an essential task is to investigate whether the protein expression levels vary among cells of different conditions. Based on the experimental results, researchers are able to verify the disease mechanism in terms of protein expression and the efficacy of treatments. Because interactions among proteins only depend on a limited number of proteins' biochemical and structural properties \citep{Terradot2004}, the manifold hypothesis is usually imposed in the analysis of protein expression levels \citep{Dorrity2020}.  

	\subsection{Delaunay Triangulation on Manifold}\label{DTM} 
	In Riemannian geometry, let $\ell_{\mathcal{M}}(\textbf{a},\textbf{b})$ denote the shortest geodesic path between two points $\textbf{a},\textbf{b}\in \mathcal{M}$,
	whose length is the geodesic distance, $d_{\mathcal{M}}(\textbf{a},\textbf{b})$, induced by the ambient space $\mathcal{R}^D$.  The manifold $\mathcal{M}$ is assumed to be geodesically convex meaning that the path $\ell_{\mathcal{M}}(\textbf{a},\textbf{b})$ uniquely exists for any $\textbf{a},\textbf{b}\in \mathcal{M}$. Let $\textbf{T}_{\textbf{a}}\mathcal{M}$ denote the tangent space of $\mathcal{M}$ at point $\textbf{a}\in\mathcal{M}$. The logarithmic map $\log_\textbf{a}: \mathcal{M} \to \textbf{T}_{\textbf{a}}\mathcal{M}$ is defined as the vector starting from $\textbf{a}$ whose length equals $d_{\mathcal{M}}(\textbf{a},\textbf{b})$ and direction is the derivative of $\ell_\mathcal{M}(\textbf{a},\textbf{b})$ at $\textbf{a}$.  
	Since  $\ell_{\mathcal{M}}(\textbf{a},\textbf{b})$ always exists uniquely for any two points $\textbf{a},\textbf{b}\in \mathcal{M}$, the domain of the logarithmic map $\log_\textbf{a}(\cdot)$ is the whole $\mathcal{M}$, for all $\textbf{a}\in \mathcal{M}$.
	We can generalize the geometric concepts from the Euclidean space to the geodesically convex Riemannian manifold $\mathcal{M}$ as follows. 
	 
	\begin{itemize}
		\item \emph{Convex set}:  
		A subset {\rm $\mathcal{C}\subset\mathcal{M}$} is convex on {\rm $\mathcal{M}$}  if for any two points {\rm $\textbf{a}, \textbf{b}\in\mathcal{C}$}, {\rm $\ell_{\mathcal{M}}(\textbf{a},\textbf{b})\subset\mathcal{C}$}.
		\item \emph{Convex hull}: For a set of points $\mathbb{Z}\subset\mathcal{M}$, the convex hull of $\mathbb{Z}$, denoted by {\rm$\mathcal{H}_\mathcal{M}(\mathbb{Z})\subset\mathcal{M}$}, is the intersection of all convex sets on $\mathcal{M}$ that contain $\mathbb{Z}$.
		\item \emph{Ball and sphere}: A ball on $\mathcal{M}$ with the center {\rm $\textbf{z}\in \mathcal{M}$} and radius $r>0$ is defined as
		$\mathcal{B}_\mathcal{M}(\textbf{z},r)=\{\textbf{z}^*\in \mathcal{M};d_{\mathcal{M}}(\textbf{z},\textbf{z}^*)\leq  r\},$
		and the corresponding sphere is defined as
		$\partial {\mathcal{B}}_\mathcal{M}(\textbf{z},r)=\{\textbf{z}^*\in \mathcal{M};d_{\mathcal{M}}(\textbf{z},\textbf{z}^*)= r\}.$
		\item \emph{Simplex and circumscribing ball}: For $k=0,\ldots,d$, the $k$-simplex of $k+1$ points {\rm$\textbf{z}_1,\ldots,\textbf{z}_{k+1}\in \mathcal{M}$} is the convex hull {\rm$\mathcal{H}_\mathcal{M}(\{\textbf{z}_1,\ldots,\textbf{z}_{k+1}\})$}, and the circumscribing ball of {\rm$\textbf{z}_1,\ldots,\textbf{z}_{k+1}$} is the smallest radius ball among all balls on $\mathcal{M}$ containing {\rm$\textbf{z}_1,\ldots,\textbf{z}_{k+1}$}. In particular, the $1$-simplex of points $\{\textbf{z}_1,\textbf{z}_2\}$ is the shortest geodesic path $\ell_{\mathcal{M}}(\textbf{z}_1,\textbf{z}_2)$.
		\item \emph{Projection}: For a set of points $\mathbb{Z}\subset\mathcal{M}$, the projection $\textbf{p}^*$ of {\rm$\textbf{p}\in \mathcal{M}$} on $\mathcal{H}_{\mathcal{M}}(\mathbb{Z})$ is defined as $\textbf{p}^*=\mathop{\arg\min}_{\textbf{q}\in\mathcal{H}_{\mathcal{M}}(\mathbb{Z})}d_{\mathcal{M}}(\textbf{p},\textbf{q})$, in the case that the right hand side exists uniquely; otherwise, we first find $\textbf{p}^\dagger=\mathop{\arg\min}_{\textbf{q}\in\mathcal{H}_{\mathcal{R}^D}(\mathbb{Z})}d_{\mathcal{R}^D}(\textbf{p},\textbf{q})$ and define  $\textbf{p}^*$ as the Euclidean projection of $\textbf{p}^\dagger$ onto $\mathcal{H}_{\mathcal{M}}(\mathbb{Z})$, which exists uniquely almost everywhere with respect to the Lebesgue measure of $\mathcal{R}^D$ \citep{Bhattacharya2003}.
	\end{itemize}
	 
	Following~\citet{Leibon2000}, the Delaunay triangulation can be defined on $\mathcal{M}$ as follows.
	
	\begin{df}\label{DT}
		For a set of $n$ points {\rm $\mathbb{Z}=\{\textbf{z}_1,\ldots,\textbf{z}_{n}\}\subset\mathcal{M}$}, the Delaunay triangulation of $\mathbb{Z}$, denoted by $\mathcal{DT}_\mathcal{M}(\mathbb{Z})$, is a mesh of $m$ $d$-simplices $\{\mathcal{S}_1,\ldots,\mathcal{S}_m\}$  satisfying:
		\begin{enumerate}
			\item For $j=1,\ldots,m$, the set of $d+1$ vertices of simplex $\mathcal{S}_j$, denoted by $\mathbb{V}(\mathcal{S}_j)$, is a subset of $\mathbb{Z}$ and is not contained in any $k$-simplex ($k=0,\ldots,d-1$) on $\mathcal{M}$.
			\item Different simplices are disjoint except on their shared boundaries.
			\item The union $\mathcal{S}_1\cup\cdots\cup\mathcal{S}_m$ is the convex hull of $\mathbb{Z}$ on $\mathcal{M}$, $\mathcal{H}_{\mathcal{M}}(\mathbb{Z})$.
			\item The empty ball property: For $j=1,\ldots,m$, the circumscribing ball of $\mathcal{S}_j$ on $\mathcal{M}$ contains no points of $\mathbb{Z}$ except on its sphere.
		\end{enumerate}
	\end{df}
	
\begin{figure}[t]
		\centering
		\begin{subfigure}{0.326\linewidth}
			\includegraphics[width=\linewidth]{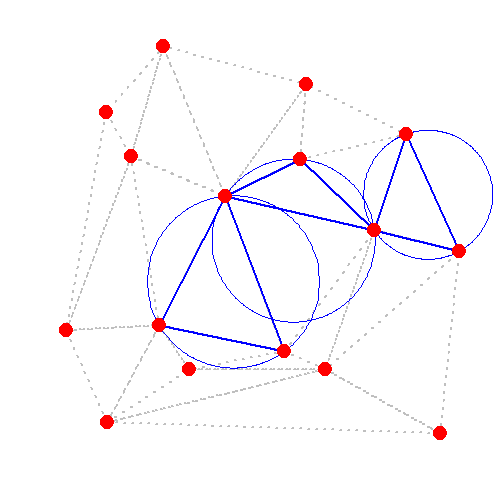}
			\caption{}
		\end{subfigure}
		\begin{subfigure}{0.326\linewidth}
			\includegraphics[width=\linewidth]{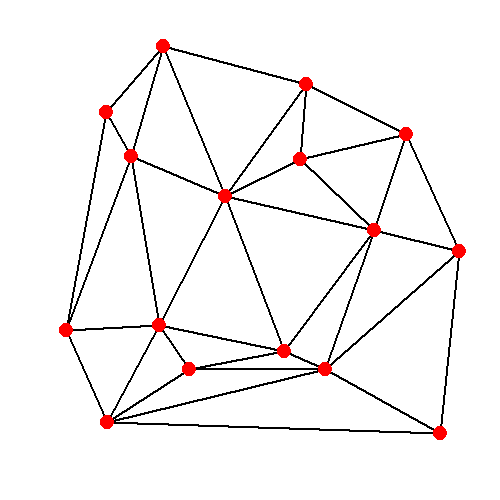}
			\caption{}
		\end{subfigure}
		\begin{subfigure}{0.326\linewidth}
			\includegraphics[width=\linewidth]{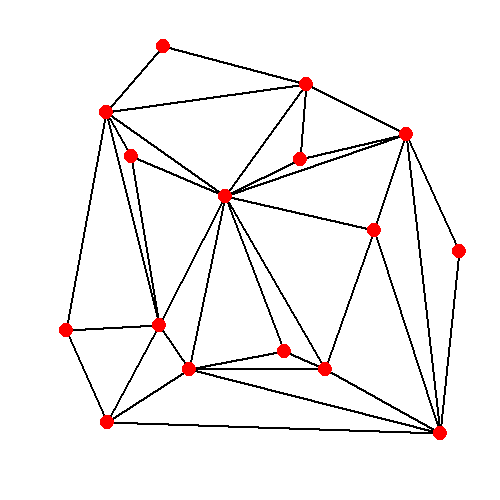}
			\caption{}
		\end{subfigure}
		\caption{(a) Graphical illustration of the empty ball property of the Delaunay triangulation with $\mathcal{M}=\mathcal{R}^2$; (b)--(c) The Delaunay triangulation versus  a random triangulation of the same $\mathbb{Z}$.}
		\label{fig:dtexample}
	\end{figure}

	We refer to simplices $\mathcal{S}_1,\ldots,\mathcal{S}_m\in \mathcal{DT}_\mathcal{M}(\mathbb{Z})$ as the Delaunay simplices. According to \citet{Leibon2000}, the Delaunay triangulation $\mathcal{DT}_\mathcal{M}(\mathbb{Z})$ is unique if and only if the pooled sample $\mathbb{Z}$ is generic, i.e., any $d+2$ points in $\mathbb{Z}$ do not lie on a sphere on $\mathcal{M}$ and any $k+2$ points do not lie in a $k$-simplex on $\mathcal{M}$ ($k=1,\ldots,d-1$). As long as both $F$ and $G$ have Lipschitz continuous densities $f$ and $g$ with respect to the measure on $\mathcal{M}$ induced by the Lebesgue measure of the ambient Euclidean space, the pooled sample $\mathbb{Z}$ is generic with probability one.
	Being the geometric dual of the Voronoi diagram, the Delaunay triangulation generates a mesh of simplices on $\mathcal{M}$ that are most regularized in shape. Figure \ref{fig:dtexample} shows the empty ball property of the Delaunay triangulation $\mathcal{DT}_{\mathcal{M}}(\mathbb{Z})$ for $\mathcal{M}=\mathcal{R}^2$, 
	and $\mathcal{DT}_{\mathcal{M}}(\mathbb{Z})$ maximizes the minimum angle in all the triangles ($2$-simplices) over all possible triangulations of $\mathbb{Z}$ \citep{Sibson1978}.
	
	\section{Delaunay Weighted Test}\label{Method} 
	
	\subsection{Delaunay Weight}\label{DWW}
	Considering the simplest case $\mathcal{M}=\mathcal{R}^d$, for each internal point $\textbf{p}\in \mathcal{H}_{\mathcal{R}^d}(\mathbb{Z})=\mathcal{S}_1\cup\cdots\cup\mathcal{S}_m$, there exists a  Delaunay simplex $\mathcal{S}(\textbf{p};\mathbb{Z})\in \mathcal{DT}_{\mathcal{R}^d}(\mathbb{Z})$ such that $\textbf{p}\in \mathcal{S}(\textbf{p};\mathbb{Z})$. We define the Delaunay weight vector for $\textbf{p}\in\mathcal{H}_{\mathcal{R}^d}(\mathbb{Z})$ with respect to $\mathbb{Z}$ as a vector $\textbf{w}(\textbf{p};\mathbb{Z})=(w_1(\textbf{p};\mathbb{Z}),\ldots,w_n(\textbf{p};\mathbb{Z}))^\dT$ such that 
	\begin{itemize}
		\item[(a)] $w_i(\textbf{p};\mathbb{Z})\in [0,1]$ if $\textbf{z}_i\in\mathbb{V}(\mathcal{S}(\textbf{p};\mathbb{Z})),$ and
		$		w_i(\textbf{p};\mathbb{Z})=0$ if $\textbf{z}_i\notin\mathbb{V}(\mathcal{S}(\textbf{p};\mathbb{Z}))$, 
		where $\mathbb{V}(\mathcal{S}(\textbf{p};\mathbb{Z}))$ denotes the vertex set of the simplex $\mathcal{S}(\textbf{p};\mathbb{Z})$;
		\item[(b)] $\sum_{i=1}^{n}w_i(\textbf{p};\mathbb{Z})=1$;
		\item[(c)] $\sum_{i=1}^{n}w_i(\textbf{p};\mathbb{Z})\cdot\textbf{z}_{i}=\textbf{p}$. 
	\end{itemize}
	According to \citet{Abadie2021}, the Delaunay weight vector $\textbf{w}(\textbf{p};\mathbb{Z})$ for $\textbf{p}\in\mathcal{H}_{\mathcal{R}^d}(\mathbb{Z})$ corresponds to the convex combination of $\mathbb{Z}$ with the smallest compound discrepancy $\sum_{i=1}^{n}w_{i}\|\textbf{p}-\textbf{z}_{i}\|^2$ among all convex combinations 
	satisfying $\textbf{p} = \sum_{i=1}^n w_i \textbf{z}_i$. That is,  $\textbf{w}(\textbf{p};\mathbb{Z})$ selects at most $d+1$ points in $\mathbb{Z}$ (those with $w_i>0$) that are convexly combined to form $\textbf{p}$ and closest to $\textbf{p}$ in the sense of the minimal compound discrepancy. 
	
	Conditions (a)--(c) only define the Delaunay weight vector for an internal point $\textbf{p}\in\mathcal{H}_{\mathcal{R}^d}(\mathbb{Z})$. For generalization to an external point $\textbf{p}\notin\mathcal{H}_{\mathcal{R}^d}(\mathbb{Z})$, we utilize the projection idea from Section 2.3 of \citet{Abadie2021}. 
	Let $\textbf{w}^{(\lambda)}(\textbf{p};\mathbb{Z})=(w^{(\lambda)}_1(\textbf{p};\mathbb{Z}),\ldots,w^{(\lambda)}_n(\textbf{p};\mathbb{Z}))^\dT$ be the solution to the optimization problem,
	$$
	\min\Bigg\|\textbf{p}-\sum_{i=1}^{n}w_{i}\textbf{z}_{i}\Bigg\|^2+\lambda \sum_{i=1}^{n}w_{i}\|\textbf{p}-\textbf{z}_{i}\|^2,\; 
	$$ 
	subject to $\textbf{w}=(w_1,\ldots,w_n)^\dT\in [0,1]^{n}$ and $\sum_{i=1}^{n}w_{i}=1$,
	where $\|\cdot\|$ denotes the Euclidean norm. The Delaunay weight vector $\textbf{w}(\textbf{p};\mathbb{Z})$ for $\textbf{p}\in\mathcal{H}_{\mathcal{R}^d}(\mathbb{Z})$ can be rewritten as
	$\textbf{w}(\textbf{p};\mathbb{Z})=\lim\limits_{\lambda\to0}\textbf{w}^{(\lambda)}(\textbf{p};\mathbb{Z})$.
	For $\textbf{p}\notin\mathcal{H}_{\mathcal{R}^d}(\mathbb{Z})$, \citet{Abadie2021} showed that $\lim\limits_{\lambda\to0}\sum_{i=1}^n{w}_i^{(\lambda)}(\textbf{p};\mathbb{Z})\textbf{z}_i=\textbf{p}^*,$ where $\textbf{p}^*$ is the projection of {$\textbf{p}$} on the convex hull $\mathcal{H}_{\mathcal{R}^d}(\mathbb{Z})$. Therefore, we have
	\begin{equation}\label{limit}\textbf{w}(\textbf{p}^*;\mathbb{Z})=\lim\limits_{\lambda\to0}\textbf{w}^{(\lambda)}(\textbf{p};\mathbb{Z}),\quad \forall \textbf{p}\in \mathcal{R}^d.
	\end{equation}
Further, we can generalize the definition of the Delaunay weight for all $\textbf{p}\in \mathcal{R}^d$ by substituting condition (c) with
	\begin{itemize}
		\item[(c$^\dagger$)]  $\sum_{i=1}^{n}w_i(\textbf{p};\mathbb{Z})\cdot\textbf{z}_{i}=\textbf{p}^*.$
	\end{itemize}

	However, when $\mathcal{M}$ is not Euclidean, condition (c$^\dagger$) is no longer applicable because it relies on the linear operation of Cartesian coordinates, which is not well-defined on $\mathcal{M}$. To generalize the Delaunay weight vector to nonlinear manifolds, we make use of the logarithmic map.
	
	\begin{df}\label{DW}
		{   For a set $\mathbb{Z}=\{\mathbf{z}_1,\ldots,\mathbf{z}_n\}\subset\mathcal{M}$, the Delaunay weight vector of {\rm$\textbf{p}\in\mathcal{M}$} is {\rm$\textbf{w}(\textbf{p};\mathbb{Z})=(w_1(\textbf{p};\mathbb{Z}),\ldots,w_n(\textbf{p};\mathbb{Z}))^\dT$} satisfying {\rm(a)}, {\rm(b)} and
			\begin{enumerate}
				\item[{\rm(c$^\ddagger$)}] {\rm$\sum_{i=1}^{n}w_i(\textbf{p};\mathbb{Z})
					\log_{\textbf{p}^*}(\textbf{z}_i)  = \mathbf{0} $,}
			\end{enumerate}
			where {\rm$\textbf{p}^*$} is the projection of {\rm$\textbf{p}$} on the convex hull $\mathcal{H}_{\mathcal{M}}(\mathbb{Z})$ and {\rm$\mathcal{S}(\textbf{p};\mathbb{Z})\in\mathcal{DT}_{\mathcal{M}}(\mathbb{Z})$} in {\rm(a)} is the Delaunay simplex containing {\rm$\textbf{p}^*$}.}
	\end{df}
	
	For the special case that $\mathcal{M}=\mathcal{R}^d$, (c$^\dagger$) and (c$^\ddagger$) are equivalent because
	\begin{equation}\label{equivalence}
		\begin{aligned}[b]
    \sum_{i=1}^{n}w_i(\textbf{p};\mathbb{Z})\cdot\overrightarrow{\textbf{p}^*\textbf{z}_{i}}=\overrightarrow{\textbf{p}^*\textbf{p}^*}
			&\Longleftrightarrow \sum_{i=1}^{n}w_i(\textbf{p};\mathbb{Z}) \log_{\textbf{p}^*}(\textbf{z}_i) = \log_{\textbf{p}^*}(\textbf{p}^*) ,
		\end{aligned}
	\end{equation}
	where $\overrightarrow{\textbf{p}^*\textbf{z}_{i}}$ denotes the vector from $\textbf{p}^*$ to $\textbf{z}_{i}$. Therefore, condition (c$^\ddagger$) is indeed a generalization of (c$^\dagger$) on a nonlinear manifold. Since the Delaunay triangulation $\mathcal{DT}_\mathcal{M}(\mathbb{Z})$ is unique for generic $\mathbb{Z}$ \citep{Leibon2000}, the Delaunay weight vector {\rm$\textbf{w}(\textbf{p};\mathbb{Z})$} is unique for $\mathbb{Z}$ and any point $\textbf{p}\in\mathcal{M}$, as shown in Section S1
	of the supplementary material.
	Based on Definition \ref{DW}, the Delaunay weight matrix of $\mathbb{Z}$ is defined as $\boldsymbol{\Gamma}_{\mathbb{Z}}=(\gamma_{ij})_{n\times n}$, with 
	\begin{equation}\label{DF:DWM}
		\gamma_{ii}=0,\quad (\gamma_{i1},\ldots,\gamma_{i,i-1},\gamma_{i,i+1},\ldots,\gamma_{in})^\dT=\textbf{w}(\textbf{z}_i;\mathbb{Z}\setminus\{\textbf{z}_i\}), \quad i=1,\ldots,n,
	\end{equation}
	where $\gamma_{ij}$ measures the geometric proximity of $\textbf{z}_j$ to $\textbf{z}_i$ and $\mathbb{Z}\setminus\{\textbf{z}_i\}$ represents the remaining set of $\mathbb{Z}$ excluding $\textbf{z}_i$. Because $\gamma_{ij}$ and $\gamma_{ji}$ are typically not the same, the Delaunay weight matrix $\boldsymbol{\Gamma}_{\mathbb{Z}}$ is a directed weighted adjacency matrix of $\mathbb{Z}$. Unlike the kernel, local regression and $k$-NN based approaches which only consider local distances to $\textbf{z}_i$, the Delaunay weight $\textbf{w}(\textbf{z}_i;\mathbb{Z}\setminus\{\textbf{z}_i\})$ evaluates the geometric proximity of the vertices of the Delaunay simplex $\mathcal{S}(\textbf{z}_i;\mathbb{Z}\setminus\{\textbf{z}_i\})$ to $\textbf{z}_i$ using both the distance and direction information. In particular, $\gamma_{ij}$ and $\gamma_{ik}$ can be different even if $\|\log_{\textbf{z}_i^*}(\textbf{z}_j)\|=\|\log_{\textbf{z}_i^*}(\textbf{z}_k)\|$. Indeed, we see from Definition \ref{DW} that $\textbf{w}(\textbf{z}_i;\mathbb{Z}\setminus\{\textbf{z}_i\})$ depends on the vectors $\log_{\textbf{z}_i^*}(\textbf{z}_j)$ rather than merely the distances $\|\log_{\textbf{z}_i^*}(\textbf{z}_j)\|$, for $j\neq i$. 
	
	\begin{figure}[t]
		\centering
		\includegraphics[width=1\linewidth]{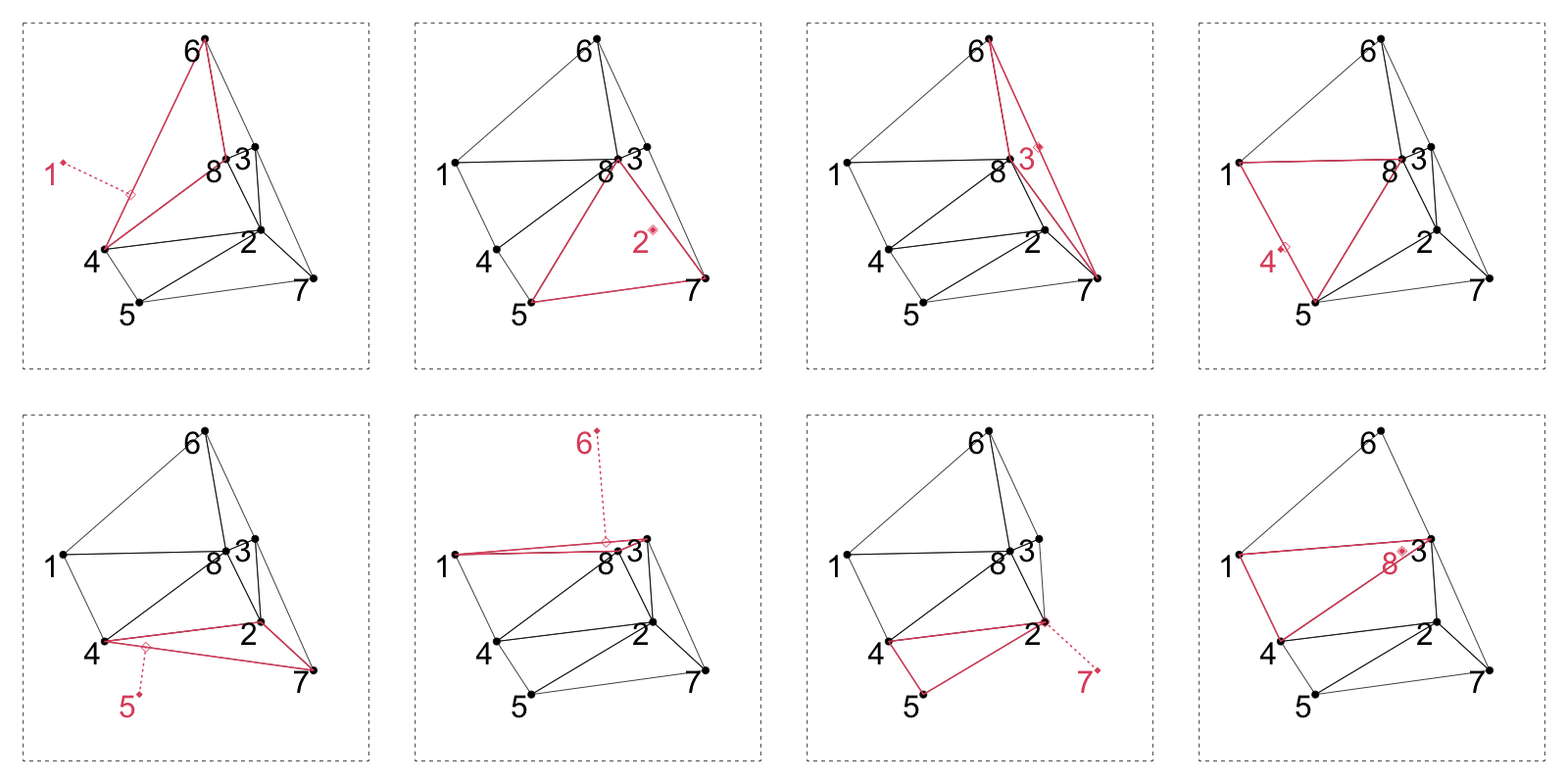}
		\caption{Graphical illustration of the Delaunay simplices (triangles) $\mathcal{S}(\textbf{z}_i;\mathbb{Z}\setminus\{\textbf{z}_i\})$ for $i=1,\ldots,4$ (top panel) and for $i=5,\ldots,8$ (bottom panel) computed from a random sample $\mathbb{Z}=\{\textbf{z}_1,\ldots,\textbf{z}_8\}$ on $\mathcal{R}^2$.}
		\label{fig:delaunay-weight}
	\end{figure}
	
	\begin{figure}[t]
		\centering
		\includegraphics[width=0.45\linewidth]{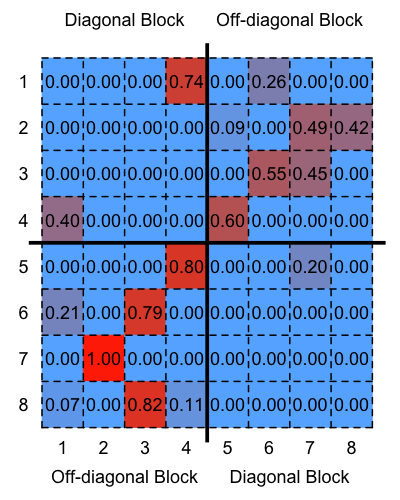}
		\caption{The Delaunay weight matrix $\boldsymbol{\Gamma}_{\mathbb{Z}}$ computed from the $\mathbb{Z}$ in Figure~\ref{fig:delaunay-weight}, with the group indicators $\delta_i=1$, for $i=1,\ldots,4$, and $\delta_i=0$ otherwise. 
			The test statistic ${T}_{\text{DW}}$ equals the average of all the Delaunay weights in the two diagonal blocks of $\boldsymbol{\Gamma}_{\mathbb{Z}}$.}
		\label{fig:delaunay-weightmatrix}
	\end{figure}
	
	To gain more insights on the Delaunay weight, we see that, when $\textbf{z}_i$ is an inner point of the Delaunay simplex $\mathcal{S}(\textbf{z}_i;\mathbb{Z}\setminus\{\textbf{z}_i\})$, the non-zero elements of $\textbf{w}(\textbf{z}_i;\mathbb{Z}\setminus\{\textbf{z}_i\})$ correspond to the points in $\mathbb{Z}\setminus\{\textbf{z}_i\}$ from all directions surrounding $\textbf{z}_i$. In contrast, under sparse or nonuniform sampling, nearest neighbors of $\textbf{z}_i$ 
	cannot cover all directions from $\textbf{z}_i$ because they tend to cluster in certain directions.
	We illustrate this point using an example with a sample $\mathbb{Z}=\{\textbf{z}_1,\ldots,\textbf{z}_8\}$ in $\mathcal{M}=\mathcal{R}^2$.  Figure \ref{fig:delaunay-weight} shows the Delaunay simplex $\mathcal{S}(\textbf{z}_i;\mathbb{Z}\setminus\{\textbf{z}_i\})$ that contains the projection of $\textbf{z}_i$ on the convex hull $\mathcal{H}_{\mathcal{M}}(\mathbb{Z}\setminus\{\textbf{z}_i\})$. Based on $\mathcal{S}(\textbf{z}_i;\mathbb{Z}\setminus\{\textbf{z}_i\})$ ($i=1,\ldots,8$), the weight vectors $\textbf{w}(\textbf{z}_i;\mathbb{Z}\setminus\{\textbf{z}_i\})$ are concatenated  to form the Delaunay weight matrix $\boldsymbol{\Gamma}_{\mathbb{Z}}$, as shown in Figure \ref{fig:delaunay-weightmatrix}. For $\textbf{z}_2\in \mathcal{H}_{\mathcal{M}}(\mathbb{Z}\setminus\{\textbf{z}_2\})$, the three nearest neighbors of $\textbf{z}_2$ are $\textbf{z}_3,\textbf{z}_7$ and $\textbf{z}_8$, which are all from the top-right direction of $\textbf{z}_2$ and the information from the bottom-left direction is not captured. In contrast, the Delaunay simplex $\mathcal{S}(\textbf{z}_2;\mathbb{Z}\setminus \{\textbf{z}_2\})$ is formed by $\textbf{z}_5,\textbf{z}_7$ and $\textbf{z}_8$, which covers all directions from $\textbf{z}_2$, and the relative directions of $\overrightarrow{\textbf{z}_2 \textbf{z}_k}$ $(k=5,7, 8)$ are used in constructing $\textbf{w}(\textbf{z}_2;\mathbb{Z}\setminus\{\textbf{z}_2\})$. Therefore, the Delaunay weight utilizes the local geometric information in a more comprehensive way than the $k$-NN.

\begin{figure}[t] 
		\centering
		\begin{subfigure}{0.32\linewidth}
			\includegraphics[width=\linewidth]{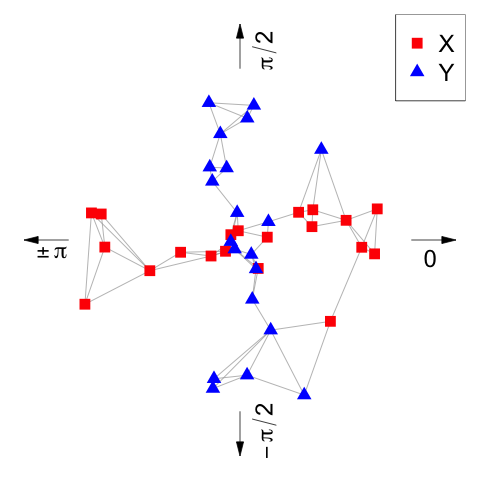}
			\caption{$k$-NN}
		\end{subfigure}
        \begin{subfigure}{0.32\linewidth}
			\includegraphics[width=\linewidth]{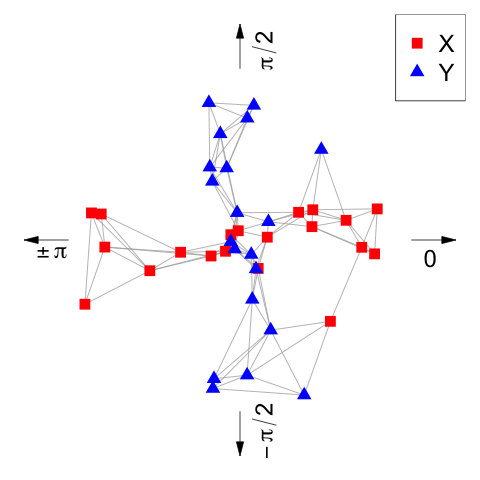}
			\caption{$k$-MST}
		\end{subfigure}
        \begin{subfigure}{0.32\linewidth}
			\includegraphics[width=\linewidth]{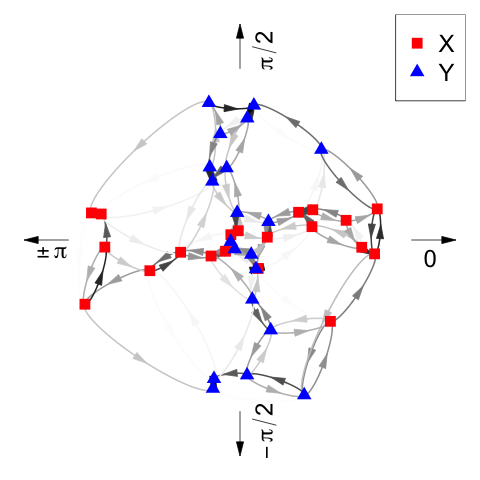}
			\caption{Delaunay weight matrix}
		\end{subfigure}
		\begin{subfigure}{0.32\linewidth}
			\includegraphics[width=\linewidth]{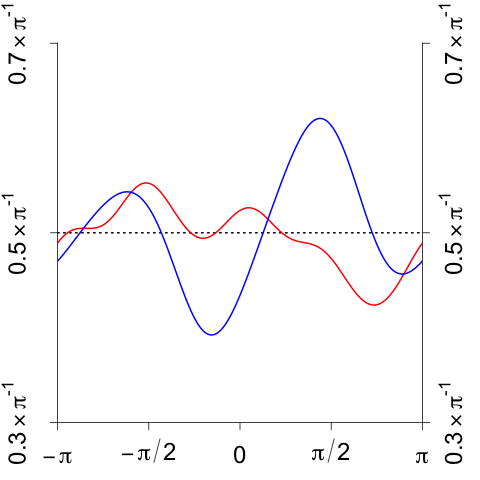}
			\caption{$k$-NN}
		\end{subfigure}
        \begin{subfigure}{0.32\linewidth}
			\includegraphics[width=\linewidth]{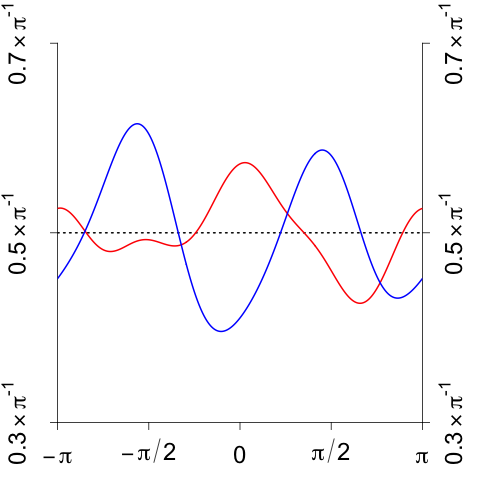}
			\caption{$k$-MST}
		\end{subfigure}
        \begin{subfigure}{0.32\linewidth}
			\includegraphics[width=\linewidth]{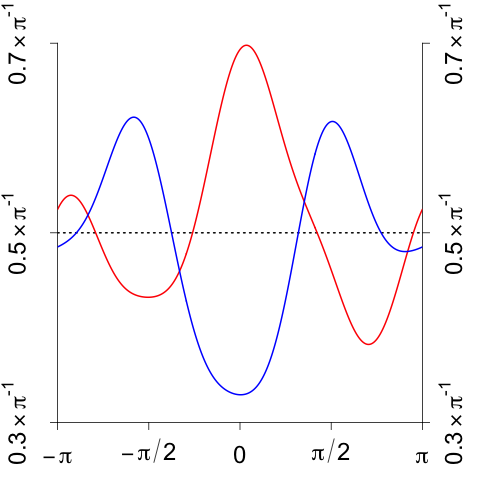}
			\caption{Delaunay weight matrix}
		\end{subfigure}
		\caption{Graphical illustration of the global advantage of the Delaunay weight matrix in capturing direction information. (a)--(c): Adjacency graphs induced by $k$-NN, $k$-MST and the Delaunay weight matrix respectively. In (c), the shade of the edge $\textbf{z}_i\to\textbf{z}_j$ indicates the magnitude of $\gamma_{ij}$. (d)--(f): The kernel density estimates (KDEs) of directions among edges pointing to $\mathbf{x}_i$'s (red) and $\mathbf{y}_i$'s  (blue) in neighborhood graphs induced by $k$-NN, $k$-MST and the Delaunay weight matrix, respectively.}\label{Directional_information}
	\end{figure}
	
	To better demonstrate the global advantage of the Delaunay weight matrix in capturing direction information, we generate a dataset $\mathbb{Z}$ in $\mathcal{R}^2$, where $\mathbb{Z}=\mathbb{X}\cup\mathbb{Y}$, $\mathbb{X}=\{\textbf{x}_1,\ldots,\textbf{x}_{20}\}$ contains 20 points from $\text{MVN}(\mathbf{0},\text{diag}(4,1))$, and $\mathbb{Y}=\{\textbf{y}_1,\ldots,\textbf{y}_{20}\}$ is obtained by rotating $\mathbb{X}$ by $\pi/2$ counterclockwise with respect to the origin. Here, we display in Figure \ref{Directional_information} (a)--(c) the adjacency graphs of $\mathbb{Z}$ induced by $k$-NN, $k$-MST \citep{Chen2017,Chen2018} and $\boldsymbol{\Gamma}_{\mathbb{Z}}$ respectively. Specifically, $k=3$ for both $k$-NN and $k$-MST, and the adjacency graph induced by $\boldsymbol{\Gamma}_{\mathbb{Z}}$ is directed with weight $\gamma_{ij}>0$ for the edge $\textbf{z}_i\to\textbf{z}_j$. Unlike $k$-NN and $k$-MST graphs, the adjacency graph induced by the Delaunay weight matrix includes more edges connecting points in the principal directions of $\mathbb{X}$ and $\mathbb{Y}$. In this example, the principal directions of $\mathbb{X}$ is along the $X$-axis (0 and $\pm \pi$), and that of $\mathbb{Y}$ is along the $Y$-axis ($\pm \pi/2$).
    As a result, the Delaunay weight matrix better captures the principal directions of $\mathbb{X}$ and $\mathbb{Y}$. To visualize this more explicitly, we show the kernel density estimate (KDE) of directions among edges pointing $\mathbf{x}_i$'s and $\mathbf{y}_i$'s (see Section S2 of the supplementary material for computing the KDEs of direction) in Figure \ref{Directional_information} (d)--(f). Specifically,  Figure \ref{Directional_information} (f) shows that among edges pointing to $\textbf{x}_i$'s, more weights are assigned to edges in directions close to the principal directions of $\mathbb{X}$. The same is true for $\mathbb{Y}$. In contrast, directions of edges pointing $\textbf{x}_i$'s in both $k$-NN and $k$-MST are less prominent in the principal directions as shown in Figure \ref{Directional_information} (d)--(e).
	
	\subsection{Test Statistic}\label{test_statistic}
	For the pooled sample $\mathbb{Z}=\{\textbf{z}_1,\ldots,\textbf{z}_n\}$ with the group indicator $\boldsymbol{\delta}=\{\delta_1,\ldots,\delta_n\}$, the $k$-NN test statistic in \citet{Schilling1986} is 
	$$\sum_{j=1}^n \sum_{i\neq j}\sum_{l=1}^kI(\textbf{z}_j\text{ is the }l\text{-th nearest neighbor of }\textbf{z}_i)\times I(\delta_i=\delta_j),$$
	which has inspired a family of nonparametric two-sample tests for detecting distribution differences via a proximity graph of the pooled sample \citep{Rosenbaum2005,Chen2017,Chen2018}. Based on the Delaunay weight $\gamma_{ij}$, we construct the test statistic,	\begin{equation}\label{Test_statistic}
		{T}_{\text{DW}}= \frac{1}{n} \sum_{j=1}^n \sum_{i\neq j} \gamma_{ij}\times I(\delta_i=\delta_j),
	\end{equation} 
	where the new geometric proximity measure $\gamma_{ij}$ accounts for the manifold structure and direction information.
	
	\subsection{Approximation of Delaunay Weight}\label{isomap} 
	In practice, the underlying manifold $\mathcal{M}$ is typically unknown, so that exact computation of the Delaunay weight matrix $\boldsymbol{\Gamma}_{\mathbb{Z}}$ via Definition~\ref{DW} (and thus $T_{\text{DW}}$) is infeasible. Because $\boldsymbol{\Gamma}_{\mathbb{Z}}$ is constructed by Delaunay weight vectors
	$\textbf{w}(\textbf{z}_i;\mathbb{Z}\setminus\{\textbf{z}_i\})$
	which only measure the geometric proximity between $\textbf{z}_i$ and the points locally surrounding $\textbf{z}_i$ (i.e., the vertices of $\mathcal{S}(\textbf{z}_i;\mathbb{Z}\setminus \{\textbf{z}_i\})$), we can utilize manifold learning techniques to approximate $\boldsymbol{\Gamma}_{\mathbb{Z}}$ by $\boldsymbol{\Gamma}_{\widetilde{\mathbb{Z}}}$, where
	$\widetilde{\mathbb{Z}}=\{\tilde{\textbf{z}}_1,\ldots,\tilde{\textbf{z}}_n\}\subset\mathcal{R}^d$ is a low-dimensional Euclidean representation of $\mathbb{Z}$ whose Delaunay weight matrix can be computed. 
	Generally speaking, any nonlinear dimension reduction technique that aims to preserve the local geometric structure of $\mathbb{Z}$ can be applied to obtain $\widetilde{\mathbb{Z}}$. {We follow the Isomap algorithm \citep{Tenenbaum2000} for its broad applicability, which requires an adjacency graph. A $k$-NN or $\epsilon$-ball graph is often used, but it cannot ensure connectivity. If the graph contains more than one connected components, then elements in $\boldsymbol{\Gamma}_{\widetilde{\mathbb{Z}}}$ corresponding to all pairs of points from different components
	must be zero, which incurs information loss and power deterioration. 
	To circumvent this issue, we propose using the union of a $k$-NN graph and the MST graph as our adjacency graph. We then estimate the geodesic distances ${d}_{\mathcal{M}}(\textbf{z}_i,\textbf{z}_j)$ by applying Dijkstra's algorithm to our adjacency graph and apply the classical MDS \citep{Kruskal1964} to obtain $\widetilde{\mathbb{Z}}\subset \mathcal{R}^{{d}}$; see Section S3
	of the supplementary material for details.} In case of unknown intrinsic dimension 
	$d$, any existing method can be applied to estimate $d$, e.g., \citet{Facco2017} as summarized in Section S3
	of the supplementary material. Our numerical experiments show that our method is robust to estimation of $d$. 
	
	To compute $\boldsymbol{\Gamma}_{\widetilde{\mathbb{Z}}}$, we note the equivalence of conditions (c$^\dagger$) and (c$^\ddagger$) in (\ref{equivalence}) for the low-dimensional Euclidean representation $\widetilde{\mathbb{Z}}$. Once we know the Delaunay simplex $\mathcal{S}(\tilde{\textbf{z}}_i;\widetilde{\mathbb{Z}}\setminus\{\tilde{\textbf{z}}_i\})$ that contains $\tilde{\textbf{z}}_i^*$, which is the projection of {\rm$\tilde{\textbf{z}}_i$} on the convex hull $\mathcal{H}_{\mathcal{R}^d}(\widetilde{\mathbb{Z}}\setminus\{\tilde{\textbf{z}}_i\})$, we can easily compute $\textbf{w}(\tilde{\textbf{z}}_i;\widetilde{\mathbb{Z}}\setminus\{\tilde{\textbf{z}}_i\})=(\tilde{\gamma}_{i1},\ldots,\tilde{\gamma}_{i,i-1},\tilde{\gamma}_{i,i+1},\ldots,\tilde{\gamma}_{in})^\dT$ by solving the linear equations,   
	\begin{equation}\label{linear}
		\begin{cases}
			\tilde{\gamma}_{ij}=0,& \forall\tilde{\textbf{z}}_j\notin\mathbb{V}(\mathcal{S}(\tilde{\textbf{z}}_i;\widetilde{\mathbb{Z}}\setminus\{\tilde{\textbf{z}}_i\})),\\
			\sum_{\tilde{\textbf{z}}_j\in\mathbb{V}(\mathcal{S}(\tilde{\textbf{z}}_i;\widetilde{\mathbb{Z}}\setminus\{\tilde{\textbf{z}}_i\}))}\tilde{\gamma}_{ij}=1,&\\ \sum_{\tilde{\textbf{z}}_j\in\mathbb{V}(\mathcal{S}(\tilde{\textbf{z}}_i;\widetilde{\mathbb{Z}}\setminus\{\tilde{\textbf{z}}_i\}))}\tilde{\gamma}_{ij}\tilde{\textbf{z}}_j=\tilde{\textbf{z}}_i^*.&\\
		\end{cases}
	\end{equation}
	The remaining difficulty in computing $\boldsymbol{\Gamma}_{\widetilde{\mathbb{Z}}}$ lies in finding the Delaunay simplex $\mathcal{S}(\tilde{\textbf{z}}_i;\widetilde{\mathbb{Z}}\setminus\{\tilde{\textbf{z}}_i\})$, for $i=1,\ldots,n$.
	
	Although the \textsf{DELAUNAYSPARSE} algorithm by \citet{Chang2020} can efficiently find $\mathcal{S}(\tilde{\textbf{z}}_i;\widetilde{\mathbb{Z}}\setminus\{\tilde{\textbf{z}}_i\})$ for	$\tilde{\textbf{z}}_i\in\mathcal{H}_{\mathcal{R}^d}(\widetilde{\mathbb{Z}}\setminus\{\tilde{\textbf{z}}_i\})$, it does not work for those $\tilde{\textbf{z}}_i\notin\mathcal{H}_{\mathcal{R}^d}(\widetilde{\mathbb{Z}}\setminus\{\tilde{\textbf{z}}_i\})$.  To tackle this difficulty, we incorporate the inverse stereographic projection to the \textsf{DELAUNAYSPARSE} algorithm and develop the stereographic projected \textsf{DELAUNAYSPARSE} algorithm to find $\mathcal{S}(\tilde{\textbf{z}}_i;\widetilde{\mathbb{Z}}\setminus\{\tilde{\textbf{z}}_i\})$, regardless of whether $\tilde{\textbf{z}}_i\in\mathcal{H}_{\mathcal{R}^d}(\widetilde{\mathbb{Z}}\setminus\{\tilde{\textbf{z}}_i\})$ or not. In summary, the stereographic projected \textsf{DELAUNAYSPARSE} algorithm first projects $\widetilde{\mathbb{Z}}\subset\mathcal{R}^d$ to $\Check{\mathbb{Z}}\subset\partial {\mathcal{B}}_{\eta r_{\text{max}}}$, then locates $\mathcal{S}(\check{\textbf{z}}_i;\Check{\mathbb{Z}}\setminus\{\check{\textbf{z}}_i\})$ in a breadth-first search manner (similar to the crystallization search in \citet{Gu2021}) and finally projects $\mathcal{S}(\check{\textbf{z}}_i;\Check{\mathbb{Z}}\setminus\{\check{\textbf{z}}_i\})$ back to $\mathcal{R}^d$ for $\mathcal{S}(\tilde{\textbf{z}}_i;\widetilde{\mathbb{Z}}\setminus\{\tilde{\textbf{z}}_i\})$. Such a procedure can be regarded as an extension of the original \textsf{DELAUNAYSPARSE} algorithm \citep{Chang2020} to the case of $\tilde{\textbf{z}}_i\notin\mathcal{H}_{\mathcal{R}^d}(\widetilde{\mathbb{Z}}\setminus\{\tilde{\textbf{z}}_i\})$. Without loss of generality, let the Euclidean representation $\widetilde{\mathbb{Z}}$ be centered at $\textbf{0}^{d}$. The stereographic projected \textsf{DELAUNAYSPARSE} algorithm is detailed as follows.
    
	\begin{figure}[t]
		\centering
		\begin{subfigure}{0.49\linewidth}
			\includegraphics[width=\linewidth]{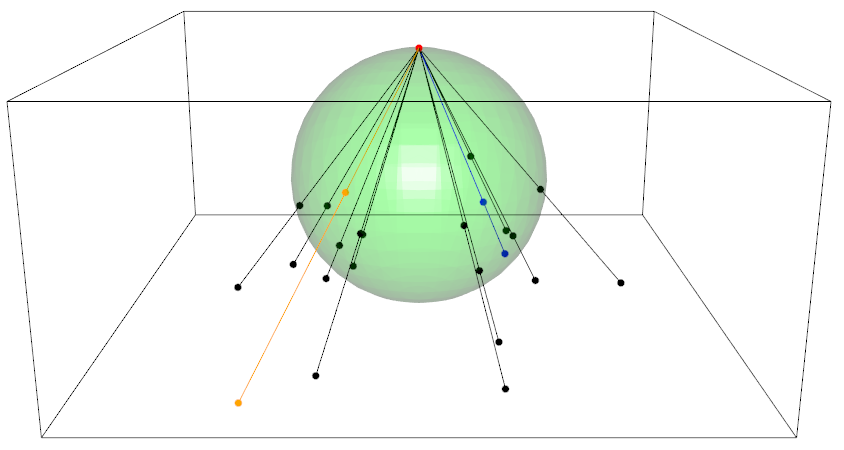}
			\caption{Inverse stereographic projection.}
		\end{subfigure}
		\begin{subfigure}{0.49\linewidth}
			\includegraphics[width=\linewidth]{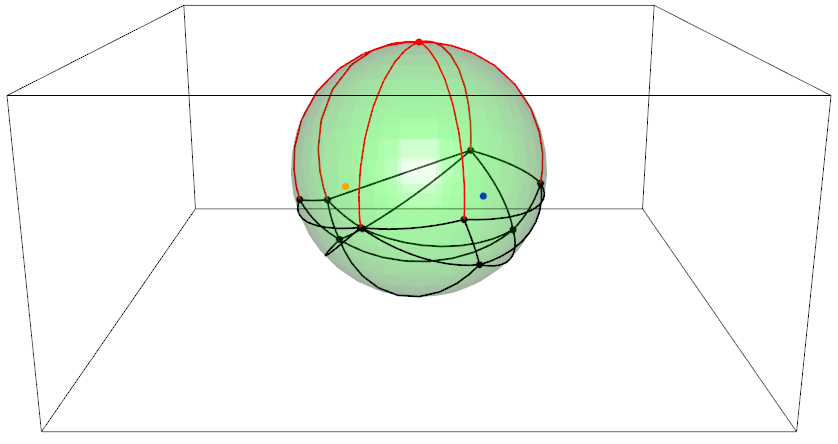}
			\caption{Delaunay triangulation on $\partial {\mathcal{B}}_{\eta r_{\text{max}}}$.}
		\end{subfigure}
		\begin{subfigure}{0.49\linewidth}
			\includegraphics[width=\linewidth]{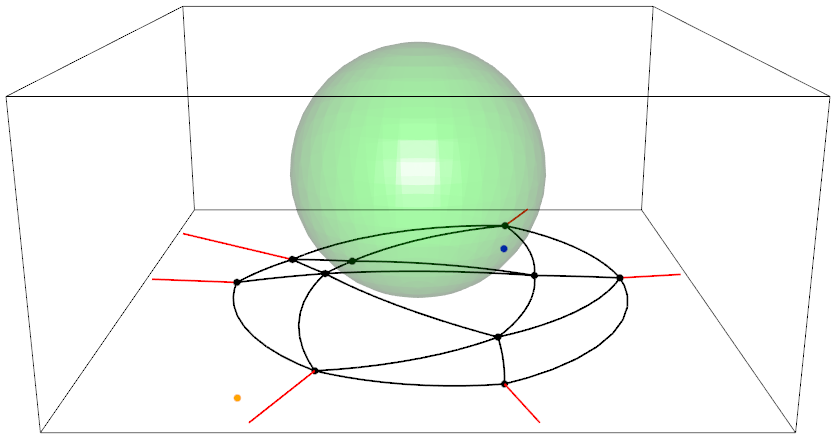}
			\caption{Projecting the Delaunay triangulation  on $\partial {\mathcal{B}}_{\eta r_{\text{max}}}$ back to $\mathcal{R}^d$ leads to an approximation of the Delaunay triangulation on $\mathcal{R}^d$.}
		\end{subfigure}
		\begin{subfigure}{0.49\linewidth}
			\includegraphics[width=\linewidth]{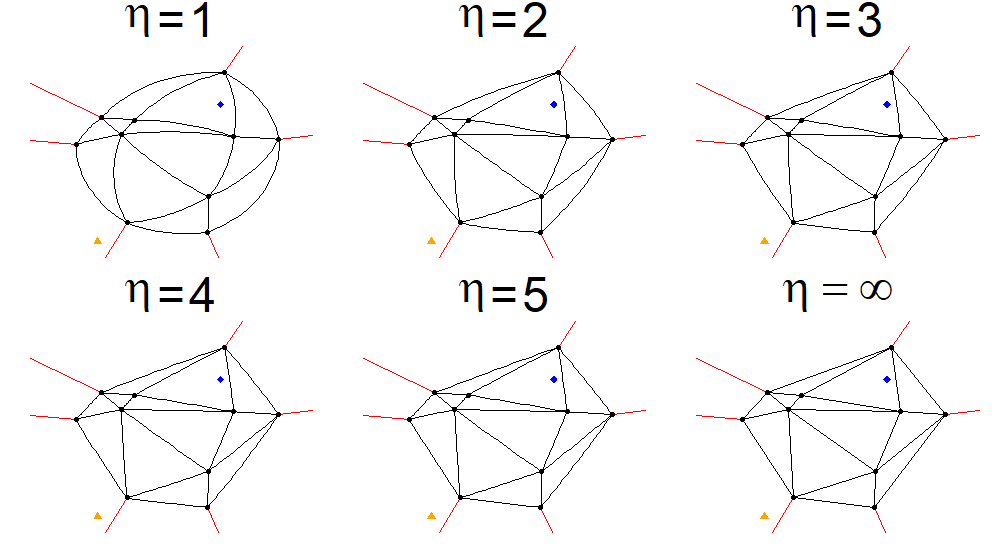}
			\caption{Approximated Delaunay triangulation approaches the true one as the scaling parameter $\eta$ grows.}
		\end{subfigure}
		\caption{Graphical illustration of the stereographic projected \textsf{DELAUNAYSPARSE} algorithm.}\label{example}
	\end{figure}
	
	\begin{itemize}
		\item[1.] We first compute $r_{\text{max}}=\max_{i=1,\ldots,n}\|\tilde{\textbf{z}}_i\|$, set a scaling parameter $\eta>0$ and apply the inverse stereographic projection,
		\begin{equation}\label{inverse}
			\phi_{\eta r_{\text{max}}}^{-1}:\mathcal{R}^{d}\to\partial {\mathcal{B}}_{\eta r_{\text{max}}} \setminus\{(\mathbf{0}_{d},\eta r_{\text{max}})^\dT\}, \quad
			\tilde{\textbf{z}}\mapsto\Bigg(\frac{\eta^2r_{\text{max}}^2}{\eta^2r_{\text{max}}^2+\|	 \tilde{\textbf{z}}\|^2}\cdot\tilde{\textbf{z}},\frac{\eta r_{\text{max}}\|\tilde{\textbf{z}}\|^2}{\eta^2r_{\text{max}}^2+\|\tilde{\textbf{z}}\|^2}\Bigg)^\dT,
		\end{equation}
		to project $\widetilde{\mathbb{Z}}$ to a $d$-sphere $\partial {\mathcal{B}}_{\eta r_{\text{max}}} =\{\textbf{z}\in \mathcal{R}^{d+1};\|\textbf{z}-(\mathbf{0}_{d},\eta r_{\text{max}}/2)^\top \|=\eta r_{\text{max}}/2\}$. Let $\check{\textbf{z}}_i=\phi_{\eta r_{\text{max}}}^{-1}(\tilde{\textbf{z}}_i)$, for $i=1,\ldots,n$, and let the reference point $\check{\textbf{z}}_0$ be $(\mathbf{0}_{d},\eta r_{\text{max}})$. We define the augmented point set as $\Check{\mathbb{Z}}=\{\check{\textbf{z}}_0,\check{\textbf{z}}_1,\ldots,\check{\textbf{z}}_n\}\subset\partial {\mathcal{B}}_{\eta r_{\text{max}}}$. A graphical illustration of obtaining $\Check{\mathbb{Z}}$ from ${\widetilde{\mathbb{Z}}}$ with $d=2$ is exhibited in Figure \ref{example} (a).
		\item[2.] As shown in Figure \ref{example} (b), for any $i=1,\ldots,n$, the union of simplices in the Delaunay triangulation $\mathcal{DT}_{\partial {\mathcal{B}}_{\eta r_{\text{max}}}} (\Check{\mathbb{Z}}\setminus\{\check{\textbf{z}}_i\})$ covers the whole sphere $\partial {\mathcal{B}}_{\eta r_{\text{max}}}$. This is due to the augmentation of $\check{\textbf{z}}_0=(\mathbf{0}_{d},\eta r_{\text{max}})$, which is the image of points at infinity on $\mathcal{R}^d$ under the inverse stereographic projection $\phi_{\eta r_{\text{max}}}^{-1}$. By projecting $\mathcal{DT}_{\partial {\mathcal{B}}_{\eta r_{\text{max}}}}  (\Check{\mathbb{Z}}\setminus\{\check{\textbf{z}}_i\}) $ back to $\mathcal{R}^d$ under the convention that the geodesic path $\ell_{\partial {\mathcal{B}}_{\eta r_{\text{max}}}}(\check{\textbf{z}}_j,\check{\textbf{z}}_0)$ (for any $j\neq i$) is mapped to the line in $\mathcal{R}^d$ from $\tilde{\textbf{z}}_j$ in the opposite direction of $\mathbf{0}_{d}$, an approximation of $\mathcal{DT}_{\mathcal{R}^d}({\widetilde{\mathbb{Z}}} \setminus \{\tilde{\textbf{z}}_i\})$ is obtained as shown in Figure \ref{example} (c). Figure \ref{example} (d) shows that such an approximation converges to $\mathcal{DT}_{\mathcal{R}^d}({\widetilde{\mathbb{Z}}} \setminus \{\tilde{\textbf{z}}_i\})$ as the scaling parameter $\eta$ tends to infinity. 
		
		\item[3.] Because $\mathcal{DT}_{\partial {\mathcal{B}}_{\eta r_{\text{max}}}} (\Check{\mathbb{Z}}\setminus\{\check{\textbf{z}}_i\})$ covers  $\partial {\mathcal{B}}_{\eta r_{\text{max}}}$, for any $i=1,\ldots,n$, there must exist a Delaunay simplex
		$\mathcal{S}^*\in \mathcal{DT}_{\partial {\mathcal{B}}_{\eta r_{\text{max}}}} (\Check{\mathbb{Z}}\setminus\{\check{\textbf{z}}_i\}) $  that contains $\check{\textbf{z}}_i$ on $\partial {\mathcal{B}}_{\eta r_{\text{max}}}$. Thus, we use the breadth-first search procedure 
		similar to the \textsf{DELAUNAYSPARSE} algorithm \citep{Chang2020} to compute the Delaunay simplex
		$\mathcal{S}^*$. Finally, if $\check{\textbf{z}}_0\notin\mathbb{V}(\mathcal{S}^*)$, $\mathcal{S}(\tilde{\textbf{z}}_i;\widetilde{\mathbb{Z}}\setminus\{\tilde{\textbf{z}}_i\})$ on $\mathcal{R}^d$ is obtained by the simplex formed by $\{\tilde{\textbf{z}}_j: \check{\textbf{z}}_j\in\mathbb{V}(\mathcal{S}^*) \}$; if $\check{\textbf{z}}_0\in\mathbb{V}(\mathcal{S}^*)$, $\mathcal{S}(\tilde{\textbf{z}}_i;\widetilde{\mathbb{Z}}\setminus\{\tilde{\textbf{z}}_i\})$ is obtained by the simplex formed by $\{\tilde{\textbf{z}}_j: \check{\textbf{z}}_j\in\mathbb{V}(\mathcal{S}^\dagger) \}$, where $\mathcal{S}^\dagger$ is the neighbor simplex of $\mathcal{S}^{*}$ opposite to $\check{\textbf{z}}_0$.
		
	\end{itemize}
	
The	complete pseudocode and other information of the stereographic projected \textsf{DELAUNAYSPARSE} algorithm are included in Section S4
	of the supplementary material. We only need to generate the local (but not full) Delaunay triangulation to find the Delaunay simplex $\mathcal{S}(\tilde{\textbf{z}}_i;\widetilde{\mathbb{Z}}\setminus\{\tilde{\textbf{z}}_i\})$, and thus the computation is not intensive. Specifically, Section S4
	of the supplementary material 
	shows that the computational complexity of the stereographic projected \textsf{DELAUNAYSPARSE} algorithm is $O(n^{2+1/d}d^2)$. As the computational complexity of solving~\eqref{linear} via Gaussian elimination is $O(nd^3)$ and the sample size $n$ is usually much larger than the intrinsic dimension $d$, we conclude that the overall computational complexity for computing the approximation $\boldsymbol{\Gamma}_{\widetilde{\mathbb{Z}}}$ is $O(n^{2+1/d}d^2)$.

The use of the inverse stereographic projection only has a small impact on the final Delaunay matrix $\boldsymbol{\Gamma}_{\widetilde{\mathbb{Z}}}$ because of its local shape-preserving property. Specifically, the inverse stereographic projection preserves the angle between the logarithmic maps and rescales the length of the logarithmic maps by a locally constant factor. This rescaling factor converges to 1 as $\eta\to\infty$. As a result, any shape in the neighborhood of $\widetilde{\textbf{z}}_i$ would be preserved with tiny distortion under the inverse stereographic projection $\phi_{\eta r_{\text{max}}}^{-1}$ when $\eta$ is large, including $\mathcal{S}(\widetilde{\textbf{z}}_i;\widetilde{\mathbb{Z}}\setminus\{\widetilde{\textbf{z}}_i\})$. Thus, the probability of returning an incorrect simplex is tiny (like in Figure \ref{example} (d)). As long as the simplex $\mathcal{S}(\widetilde{\textbf{z}}_i;\widetilde{\mathbb{Z}}\setminus\{\widetilde{\textbf{z}}_i\})$ is correctly found, the Delaunay weight can be correctly calculated using both $\widetilde{\textbf{z}}_i$ and $\mathcal{S}(\widetilde{\textbf{z}}_i;\widetilde{\mathbb{Z}}\setminus\{\widetilde{\textbf{z}}_i\})$ on $\mathcal{R}^d$. To empirically evaluate the impact of the stereographic projection parameter $\eta$ on the final Delaunay weight matrix, we perform an experiment as detailed in Section S4 of the supplementary material. It is shown that the Delaunay weight matrix $\boldsymbol{\Gamma}_{\widetilde{\mathbb{Z}},\eta}$ only differs due to unavoidable floating-point errors of the computer when $\eta\geq 15$. Thus, we suggest using $\eta=15$ in practice. 
	
	\subsection{Permutation Test}\label{permuation_test}
	
	The Delaunay weighted test statistic in \eqref{Test_statistic}, ${T}_{\text{DW}}= \sum_{j=1}^n \sum_{i\neq j} \gamma_{ij}\times I(\delta_i=\delta_j)/n$, aims to detect differences between the underlying distributions $F$ and $G$ via the deviation of $T_{\text{DW}}$ from its expectation under the null. Figure \ref{fig:delaunay-weightmatrix} gives an illustrative example of the Delaunay weight matrix $\boldsymbol{\Gamma}_{\mathbb{Z}}$ with group indicators $\delta_i=1$, for $i=1,\ldots,4$, and $\delta_i=0$ otherwise. If we partition $\Gamma_\mathbb{Z}$ at the 4-th row and column, ${T}_{\text{DW}}$ equals the average of all Delaunay weights in the two diagonal blocks of $\boldsymbol{\Gamma}_{\mathbb{Z}}$. Generally, off-diagonal elements in $\boldsymbol{\Gamma}_{\mathbb{Z}}$ are identically distributed under the null. In contrast, under $H_1: F\neq G$, more $\textbf{z}_i$'s in $\mathbb{X}$ are located in the region where the density of $F$ is higher than that of $G$. For these $\textbf{z}_i$'s in $\mathbb{X}$, the expected proportion of data points from  $\mathbb{X}$ in $\mathbb{V}(\mathcal{S}(\textbf{z}_i;\mathbb{Z}\setminus\{\textbf{z}_i\}))$ is larger than that in $\mathbb{Z}\setminus\{\textbf{z}_i\}$. Similar arguments also hold for $\mathbb{Y}$. Thus, ${T}_{\text{DW}}$ is stochastically larger under the alternative than under the null. 
	
	The exact distribution of $T_{\text{DW}}$ under the null 
	is complex and difficult to derive, and thus we implement a permutation procedure to compute the $p$-value as follows.
	\begin{enumerate}
		\item[1.] Compute the observed value of the test statistic $T_{\text{DW}}$ via (\ref{Test_statistic}).
		\item[2.] For $b=1,\ldots,B$, we randomly generate a permutation of group indicators $\boldsymbol{\delta}^{(b)}=\{\delta_1^{(b)},\ldots,\delta_n^{(b)}\}$, where $\sum_{i=1}^n\delta_i^{(b)}=n_1$ and $\sum_{i=1}^n(1-\delta_i^{(b)})=n_0$, and compute 
		$T^{(b)}_{\text{DW}}=\sum_{j=1}^n \sum_{i\neq j}\gamma_{ij}\times I(\delta_i^{(b)}=\delta_j^{(b)})/n$ under $\boldsymbol{\delta}^{(b)}$.
		\item[3.] Compute 
		$
		p{\textrm{-value}} = \{\sum_{b=1}^{B}I(T_{\text{DW}}\leq T^{(b)}_{\text{DW}})+1\}/(B+1),
		$
		where 
		$1$ is added in both the numerator and denominator to calibrate that under $H_0$, $\text{Pr} (p\text{-value}\leq \alpha)\leq \alpha$ for all $\alpha\in [0,1]$. Indeed, under $H_0$, $T_{\text{DW}},T_{\text{DW}}^{(1)},\ldots,T_{\text{DW}}^{(B)}$ are i.i.d.~given the pooled sample $\mathbb{Z}$ and sample sizes $(n_1,n_0)$. Therefore, the $p\text{-value}$ is uniformly distributed on $\{1/(B+1),2/(B+1),\ldots,1\}$, which converges to $\text{Unif}[0, 1]$ as $B\to\infty$.
	\end{enumerate}
	Under the unknown manifold setting where $\Gamma_{\mathbb{Z}}$ cannot be computed exactly, we replace it by $\Gamma_{\widetilde{\mathbb{Z}}}$ for approximate inference. 
	Unlike most of the existing tests based on pairwise distances only, the Delaunay weight takes the relative directions among data points into account, and thus our test gains more power in detecting the direction difference between $F$ and $G$.
	
	\section{Theoretical Analysis}\label{Theory}

	For convenience, we write $\check{\gamma}_{ij}=\gamma_{ij}+\gamma_{ji}$, for all $i\neq j$, and we use 
	the subscript ``${H_k}$" in $\text{E}_{H_k}$ and
$\text{Var}_{H_k}$ to denote the condition that $H_k$ is true, for $k=0,1$.
	To derive asymptotic properties of the test statistic $T_{\text{DW}}$, it is common to deduce the distribution of the unconditional standardized statistic. 
	However, since $T_{\text{DW}}$ is a function of both $\boldsymbol{\delta}$ and the Delaunay weight matrix $\boldsymbol{\Gamma}_{\mathbb{Z}}$, it is difficult to obtain an explicit expression of its variance under the null due to the discrete nature of the Delaunay triangulation. Therefore, we first provide the expectation and variance of the test statistic conditional on $\mathbb{Z}$, where the proximity graph based on $\boldsymbol{\Gamma}_{\mathbb{Z}}$ becomes nonrandom as in \citet{Chen2017} and \citet{Chen2018}. 
	
	\begin{thm}
		\label{thm:Mean_Var}
		Let {\rm $\mathbb{Z} = \{ \textbf{z}_1,\ldots,\textbf{z}_n \} $} be a set of generic points on a $d$-dimensional geodesically convex Riemannian manifold  {\rm$\mathcal{M}\subset\mathcal{R}^D$}. When $H_0$ is true, we have
        {\rm $$\text{E}_{H_0}(T_{\text{DW}}|\mathbb{Z})=\frac{n_1(n_1-1)+n_0(n_0-1)}{n(n-1)}$$}
		and {\rm $$
			\text{Var}_{H_0}(T_{\text{DW}}|\mathbb{Z})=
			\frac{n_1(n_1-1)n_0(n_0-1)}{3n^2}V_{0,\mathbb{Z}}    +\frac{n_1n_0(n-2)}{3n^2}V_{1,\mathbb{Z}}+
			\frac{n_1n_0}{n^2}V_{2,\mathbb{Z}}-\frac{4n_1^2n_0^2}{(n-1)^2n^2},
			$$}
		where
		{\rm	$$\begin{aligned}[b]
				V_{0,\mathbb{Z}}&=\frac{\sum_{1\leq i<j<k<l\leq n}(\check{\gamma}_{ij}\check{\gamma}_{kl}+\check{\gamma}_{ik}\check{\gamma}_{jl}+\check{\gamma}_{il}\check{\gamma}_{jk})}{n(n-1)(n-2)(n-3)/24},\\
				V_{1,\mathbb{Z}}&=\frac{\sum_{1\leq i<j<k\leq n}(\check{\gamma}_{ij}\check{\gamma}_{ik}+\check{\gamma}_{ij}\check{\gamma}_{jk}+\check{\gamma}_{ik}\check{\gamma}_{jk})}{n(n-1)(n-2)/6},\\
				V_{2,\mathbb{Z}}&=\frac{\sum_{1\leq i<j\leq n}\check{\gamma}_{ij}^2}{n(n-1)/2},				
			\end{aligned}$$}
		are average values of $\{(\check{\gamma}_{ij}\check{\gamma}_{kl}+\check{\gamma}_{ik}\check{\gamma}_{jl}+\check{\gamma}_{il}\check{\gamma}_{jk})|1\leq i<j<k<l\leq n\}$,  $\{(\check{\gamma}_{ij}\check{\gamma}_{ik}+\check{\gamma}_{ij}\check{\gamma}_{jk}+\check{\gamma}_{ik}\check{\gamma}_{jk})|1\leq i<j<k\leq n\}$ and 
		$\{\check{\gamma}_{ij}^2|1\leq i<j\leq n\}$,  respectively.
	\end{thm}		
	
	The proof of Theorem \ref{thm:Mean_Var} is 
	based on the fact that $\boldsymbol{\delta}$ is uniformly distributed on all $\binom{n }{n_1}$ realizations containing $n_1$ ones and $n_0$ zeros  conditional on $\mathbb{Z}$ under $H_0$, as detailed in Section S5 
	of the supplementary material. Given the finite-sample conditional mean $\text{E}_{H_0}(T_{\text{DW}}|\mathbb{Z})$ and conditional variance $\text{Var}_{H_0}(T_{\text{DW}}|\mathbb{Z})$, we deduce the asymptotic null distribution of the standardized statistic of $T_{\text{DW}}$. 
	
	\begin{thm}\label{thm:null1}
		Assume that {\rm(i)} $\lim\limits_{n\to\infty}n_1/n=\kappa\in(0,1)$; {\rm(ii)} $F=G$ and they have Lipschitz continuous densities $f=g$ with respect to the measure on $\mathcal{M}$ induced by the Lebesgue measure on the ambient Euclidean space; and {\rm(iii)} there exists $\nu<9/4$ such that $\sum_{i=1}^n|\mathcal{N}_i|^4=o(n^{\nu})$, with {\rm$\mathcal{N}_i=\cup_{\mathcal{S}\in\mathcal{DT}_\mathcal{M}(\mathbb{Z}):\textbf{z}_i\in\mathcal{S}}\mathbb{V}(\mathcal{S})$}. As $n\to \infty$, we have
		{\rm \begin{equation}\label{asy_null}
				\frac{T_{\text{DW}}-\text{E}_{H_0}(T_{\text{DW}}|\mathbb{Z})}{\sqrt{\text{Var}_{H_0}(T_{\text{DW}}|\mathbb{Z})}}{\displaystyle\mathop{\longrightarrow}^{D}} N(0,1),		\end{equation}}
		where ${\displaystyle\mathop{\longrightarrow}^{D}}$ denotes convergence in distribution.
	\end{thm}
	
	The proof of Theorem \ref{thm:null1} is given in Section S6
	of the supplementary material. In Theorem \ref{thm:null1}, conditions (i) and (ii) are standard and mild. 
	To gain insight into condition (iii), we define a hub $\textbf{z}_i$
	as a point with a degree much larger than the average degree in the $\mathcal{DT}_\mathcal{M}(\mathbb{Z})$, which only occurs under the rare case that many data points are located on the sphere  $\partial {\mathcal{B}}_\mathcal{M}(\textbf{z}_i,r)$ for some $r>0$.
	Condition (iii) regularizes the Delaunay triangulation $\mathcal{DT}_\mathcal{M}(\mathbb{Z})$, which is satisfied if no hub exists in $\mathcal{DT}_\mathcal{M}(\mathbb{Z})$.  Empirical evidence of Theorem~\ref{thm:null1} is provided in Section S8
	of the supplementary material.
	
	The consistency of the Delaunay weighted test with the permutation procedure can be shown as follows.
	
	\begin{thm}\label{thm:consistency}
		Under the same conditions as in Theorem~\ref{thm:null1} except that $F=G$, the Delaunay weighted test 
		using {\rm$T_{\text{DW}}$} is consistent against all alternatives. That is, for any significance level $\alpha\in(0,1)$ and $F\neq G$, the Delaunay weighted test with the permutation procedure 
		rejects the null $H_0:F=G$ with asymptotic probability one.
	\end{thm}
	
	The proof of Theorem \ref{thm:consistency} is given in Section S7 
	of the supplementary material, for which the main idea is adopted from \citet{Henze1999}. In particular, we show that $T_{\text{DW}}-\text{E}_{H_0}(T_{\text{DW}}|\mathbb{Z})$ converges to a positive value and $\text{Var}_{H_0}(T_{\text{DW}}|\mathbb{Z})$ converges to zero a.s.,~as $n\to\infty$. Although the test statistic $T_{\text{DW}}$ is constructed based on $\boldsymbol{\Gamma}_{\widetilde{\mathbb{Z}}}$ instead of $\boldsymbol{\Gamma}_{ \mathbb{Z}}$ in practice, the consistency of the test is preserved as long as the mapping from $\mathbb{Z}$ to $\widetilde{\mathbb{Z}}$ is one-to-one, a standard assumption in manifold learning.
	
	\section{Experiments}\label{experiments}
	To illustrate the empirical performance of the Delaunay weighted test, we conduct extensive experiments on both simulated and real data. For comparison, we also implement several existing nonparametric tests, including the $k$-NN test \citep{Hall2002}, the $e$-distance test \citep{Szekely2004}, the covariance test \citep{Cai2013b}, the $k$-MST test \citep{Chen2017}, the regression test \citep{Kim2019} and the kernel test \citep{Gretton2012,Cheng2024}. Among these methods, the (local) regression test with the $k$-NN regression and the kernel test are adaptive to the intrinsic dimension of the underlying manifold, and thus we adopt the $k$-NN regression with $k=n^{2/(2+d)}$ as suggested by \citet{Kim2019} in the regression test. 
	Following the suggestion that the bandwidth should scale with $\sqrt{D}$ and $n^{-1/(d+2\beta)}$  \citep{Cheng2024}, we choose the bandwidth $s\cdot n^{-1/(d+2\beta)}$ for the kernel test, where $s^2$ is the sum of variances of $\mathbb{Z}$ and $\beta =1 $ is the H\"{o}lder exponent.  As the Delaunay weight $\textbf{w}(\textbf{z}_i;\mathbb{Z}\setminus\{\textbf{z}_i\})$, for each $\textbf{z}_i$, contains at most $d+1$ nonzero components, we use $k=d+1$ in the $k$-NN tests and $k$-MST test for fair comparison. The covariance test is consistent only if the covariances of $F$ and $G$ differ, while it is of interest to compare its performance under general cases. 
	For the standard permutation procedure, we set the number of permutations as $B=1000$. 
	
	\subsection{Scenario 1: Euclidean data}\label{Sim1}
	
	To validate the size of the Delaunay weighted test on Euclidean data, i.e., $\mathcal{M}=\mathcal{R}^D$, we generate 1000 Gaussian datasets under the null $H_0$. In each dataset, $\mathbb{X}$ and $\mathbb{Y}$ respectively contain $n_1$ and $n_0$ i.i.d.~samples from the multivariate normal distribution $\text{MVN}(\mathbf{0}_d,\textbf{I}_{d\times d})$, where $\textbf{I}_{d\times d}$ is the $d\times d$ identity matrix with $d=D$. We consider different dimensions $d\in \{20,50\}$, sample sizes $(n_1,n_0)\in\{(50,50),(100,50),(100,100)\}$ and significance levels $\alpha=0.01,0.05$ and $0.10$. Table \ref{Euclidean_null} shows the rejection proportions of the Delaunay weighted test under different cases.
    In general, the Delaunay weighted test is able to control the type I error rate at the target level.

    \renewcommand{\arraystretch}{0.9}
    \begin{table}[t]
        \centering
        \caption{Rejection proportions (\%) of the Delaunay weighted test on the Euclidean data under the null and different significance levels $\alpha=0.01,0.05,0.10$.}
        \label{Euclidean_null}
        \resizebox{0.9\columnwidth}{!}{\begin{tabular}{ccccccccccccc}
             \toprule
             &\multicolumn{3}{c}{$n_1=50,n_0=50$}&&\multicolumn{3}{c}{$n_1=100,n_0=50$}&&\multicolumn{3}{c}{$n_1=100,n_0=100$}\\\cline{2-4}\cline{6-8}\cline{10-12}
$d$&$\alpha=0.01$&$0.05$&$0.10$&&$0.01$&$0.05$&$0.10$&&$0.01$&$0.05$&$0.10$\\
             \midrule
             $20$&1.4&4.6&8.7&&1.0&4.7&10.3&&0.5&4.8&10.5\\
$50$&1.0&4.2&9.5&&0.6&4.5&10.1&&1.0&4.7&9.9\\

             \bottomrule
        \end{tabular}}
    \end{table}
	
	We also investigate the power of the Delaunay weighted test in distinguishing differences between distributions $F$ and $G$ on 1000 Gaussian datasets. We consider both the location alternative and the direction alternative, where the former imposes the difference in location parameters (mean or median) of distributions $F$ and $G$, and the latter sets the same location parameter for distributions $F$ and $G$ but their covariance matrices have different principal directions (directions with large variance). Under the location alternative,  $\mathbb{X}$ contains $n_1$ independent samples $\textbf{x}_1,\ldots,\textbf{x}_{n_1}$ from $F=\text{MVN}(\mathbf{0}_d,\textbf{I}_{d\times d})$, while $\textbf{y}_1,\ldots,\textbf{y}_{n_0}$ in $\mathbb{Y}$ are generated from $G=\text{MVN}(\boldsymbol{\Delta}_d,\textbf{I}_{d\times d})$, where $\boldsymbol{\Delta}_d$ is an independent variable uniformly distributed on $\partial {\mathcal{B}}_{\mathcal{R}^d}(\mathbf{0},0.8)$ for $d=20$ and $\partial {\mathcal{B}}_{\mathcal{R}^d}(\mathbf{0},1)$ for $d=50$. Under the direction alternative, both samples $\mathbb{X}$ and $\mathbb{Y}$ are generated from zero-mean Gaussian distributions $F=\text{MVN}(\mathbf{0}_d,\boldsymbol{\Sigma}_{\textbf{x}})$ and $G=\text{MVN}(\mathbf{0}_d,\boldsymbol{\Sigma}_{\textbf{y}})$, with covariance matrices, 
	$$\boldsymbol{\Sigma}_{\textbf{x}} =\begin{pmatrix}
		1.25^2 \textbf{I} & \textbf{0} \\\textbf{0} &\textbf{I} \\
	\end{pmatrix}  \mathrm{~~and~~}
	\boldsymbol{\Sigma}_{\textbf{y}} =\begin{pmatrix}
		\textbf{I} & \textbf{0} \\\textbf{0} &1.25^2 \textbf{I} \\
	\end{pmatrix},
	$$
    where all submatrices $\textbf{I}$ and $\textbf{0}$ are of dimensions $(d/2)\times (d/2)$. 

    \renewcommand{\arraystretch}{0.9}
    \begin{table}[t]
        \centering
        \caption{Rejection proportions (\%) of different two-sample tests on the Euclidean data under the location alternative and different significance levels $\alpha=0.01,0.05,0.10$, where the best performance is highlighted in boldface.}
        \label{Euclidean_location}
        \resizebox{\columnwidth}{!}{\begin{tabular}{llrrrrrrrrrrrr}
             \toprule
             &&\multicolumn{3}{c}{$n_1=50,n_0=50$}&&\multicolumn{3}{c}{$n_1=100,n_0=50$}&&\multicolumn{3}{c}{$n_1=100,n_0=100$}\\\cline{3-5}\cline{7-9}\cline{11-13}
             $d$& Tests&$\alpha =0.01$&$0.05$&$0.10$&&$0.01$&$0.05$&$0.10$&&$0.01$&$0.05$&$0.10$\\
             \midrule
             \multirow{7}*{$20$}&Delaunay weighted&11.9&29.5&42.9&&14.2&33.9&47.3&&33.7&58.2&69.0\\
&$k$-NN&6.2&21.2&33.3&&7.5&27.3&42.5&&18.6&45.6&61.0\\
&$k$-MST&19.2&31.9&41.8&&27.2&44.4&54.7&&49.7&66.6&73.7\\
&Kernel&27.1&52.2&65.9&&40.3&66.9&79.4&&71.9&87.5&92.9\\
&$e$-distance&\textbf{38.3}&\textbf{64.6}&\textbf{76.2}&&\textbf{57.8}&\textbf{81.5}&\textbf{88.7}&&\textbf{86.5}&\textbf{95.9}&\textbf{98.2}\\
&Covariance&0.9&5.3&10.5&&0.9&5.4&13.2&&1.3&7.3&13.6\\
&Regression&5.5&17.4&28.1&&6.2&18.4&33.9&&15.8&36.8&50.6\\\midrule
\multirow{7}*{$50$}&Delaunay weighted&17.0&37.6&50.4&&22.6&44.6&57.8&&40.5&65.9&77.0\\
&$k$-NN&1.9&12.4&25.2&&2.5&16.9&33.3&&4.3&29.7&50.3\\
&$k$-MST&5.6&13.5&21.1&&33.1&54.5&65.6&&63.0&81.6&86.8\\
&Kernel&39.0&64.9&74.7&&59.5&83.7&91.4&&88.1&97.0&\textbf{99.2}\\
&$e$-distance&\textbf{43.8}&\textbf{67.9}&\textbf{78.3}&&\textbf{66.0}&\textbf{87.4}&\textbf{93.6}&&\textbf{90.8}&\textbf{98.9}&\textbf{99.2}\\
&Covariance&1.1&4.9&11.2&&1.2&6.2&11.5&&0.8&5.4&9.7\\
&Regression&2.5&13.0&21.8&&1.9&10.5&20.2&&6.5&22.4&36.5\\
             \bottomrule
        \end{tabular}}
    \end{table}

    Table \ref{Euclidean_location} presents the power of the Delaunay weighted test in comparison with other methods under the location alternative. For both $d=20$ and $d=50$, the power of the Delaunay weighted test tends to $1$ as the total sample size $n_1+n_0$ grows, demonstrating its ability to distinguish the location difference of distributions. Compared with the existing ones, our test outperforms the $k$-NN test and the regression test and it yields comparable power with the $k$-MST test, while the kernel test and the $e$-distance test possess the highest power. However, the power difference between these two tests and our Delaunay weighted test becomes smaller for $d=50$.
    As expected, 
	the covariance test is powerless, because it is not consistent in distinguishing the location difference.

\renewcommand{\arraystretch}{0.9}
    \begin{table}[t]
        \centering
        \caption{Rejection proportions (\%) of different two-sample tests on the Euclidean data under the direction alternative and different significance levels $\alpha=0.01,0.05,0.10$, where the best performance is highlighted in boldface.}
        \label{Euclidean_direction}
        \resizebox{\columnwidth}{!}{\begin{tabular}{llrrrrrrrrrrrr}
             \toprule
             &&\multicolumn{3}{c}{$n_1=50,n_0=50$}&&\multicolumn{3}{c}{$n_1=100,n_0=50$}&&\multicolumn{3}{c}{$n_1=100,n_0=100$}\\\cline{3-5}\cline{7-9}\cline{11-13}
             $d$& {Tests}&$\alpha=0.01$&$0.05$&$0.10$&&$0.01$&$0.05$&$0.10$&&$0.01$&$0.05$&$0.10$\\
             \midrule
             \multirow{7}*{$20$}&Delaunay weighted&\textbf{12.1}&\textbf{31.2}&\textbf{45.7}&&\textbf{21.1}&45.0&59.6&&\textbf{48.8}&\textbf{74.0}&\textbf{85.6}\\
&$k$-NN&1.5&7.5&13.5&&1.8&8.5&16.2&&3.4&11.7&21.7\\
&$k$-MST&1.8&6.7&13.0&&6.4&14.2&20.7&&14.6&29.4&41.0\\
&Kernel&3.6&13.4&23.5&&6.5&19.6&30.9&&11.2&34.8&50.9\\
&$e$-distance&1.2&6.2&12.1&&1.6&8.9&15.0&&1.9&7.8&15.6\\
&Covariance&4.6&18.8&33.5&&18.9&\textbf{46.1}&\textbf{61.8}&&25.0&61.7&79.9\\
&Regression&1.7&7.5&14.2&&1.3&7.7&14.7&&3.5&13.0&22.7\\
\midrule
\multirow{7}*{$50$}&Delaunay weighted&\textbf{15.6}&\textbf{34.7}&\textbf{49.2}&\textbf{}&\textbf{31.6}&\textbf{56.6}&\textbf{73.0}&\textbf{}&\textbf{61.7}&\textbf{82.7}&\textbf{91.4}\\
&$k$-NN&1.3&5.5&10.6&&0.7&5.7&10.7&&0.6&5.2&14.8\\
&$k$-MST&3.3&8.4&12.9&&1.3&4.5&9.1&&1.5&6.5&12.3\\
&Kernel&2.1&9.8&18.3&&3.8&12.9&23.9&&5.9&18.8&32.4\\
&$e$-distance&0.6&5.4&11.3&&2.0&6.8&11.1&&1.2&6.2&10.8\\
&Covariance&2.7&12.9&24.5&&17.9&45.5&62.0&&19.1&52.2&68.2\\
&Regression&0.8&5.9&10.4&&1.3&5.0&10.7&&3.1&8.6&16.7\\
             \bottomrule
        \end{tabular}}
    \end{table}

    Table \ref{Euclidean_direction} presents the power of the Delaunay weighted test in comparison with other methods under the direction alternative. Our method outperforms all the tests in comparison, except that the covariance test has comparable power when $d=20$. Under the direction alternative, our test, which takes the direction information into account, is more powerful than those tests that only use the distance information. The advantage of our Delaunay weighted test over the others further amplifies when the dimension $d$ grows to $50$, because the information contained in the Euclidean distance diminishes as $d$ grows.

		\subsection{Scenario 2: Manifold Data}\label{MNIST}

        \subsubsection{Size and Power}\label{Size_Power}
		
		To validate the effectiveness of the Delaunay weighted test on high-dimensional data that lie on a low-dimensional manifold, we generate synthetic datasets on a $d$-dimensional manifold $\mathcal{M}$ embedded in $\mathcal{R}^D$ with $d=10$ and $D=100$. Specifically, the $d$-dimensional manifold $\mathcal{M}$ is obtained via the parametric transformation 
        $$\varphi:\mathcal{R}^d\mapsto\mathcal{R}^D,\quad \varphi(t_1,\ldots,t_{d})=\textbf{Q}\begin{pmatrix}
            t_1\cdot \cos(\pi t_1)\\t_1\cdot \sin(\pi t_1)\cdot \text{sign}(t_1)\\
            \vdots\\
            t_{d/2}\cdot \cos(\pi t_{d/2})\\t_{d/2}\cdot \sin(\pi t_{d/2}) \text{sign}(t_{d/2})\\
            t_{d/2+1}\\\vdots\\t_{d}
        \end{pmatrix},$$
        where $\textbf{Q}$ is a random $D \times (3d/2)$ matrix with orthonormal columns. Under this model, the geodesic distance between $\textbf{z}_1=\varphi(t_{11},\ldots,t_{1d})$ and $\textbf{z}_2=\varphi(t_{21},\ldots,t_{2d})$ is $$d_{\mathcal{M}}(\textbf{z}_1,\textbf{z}_2)=\sqrt{\sum_{i=1}^{d/2}\left\{\int_{t_{1i}}^{t_{2i}}\sqrt{1+\pi^2 s^2}ds\right\}^2+\sum_{i=d/2+1}^{d}(t_{2i}-t_{1i})^2}.$$
        
        To validate the size and power of the Delaunay weighted test on manifold data, we randomly generate 1000 synthetic datasets with sample sizes $(n_1,n_0)\in \{(250,250),(500,500)\}$. For $i=1,\ldots,n_1$, we generate $\textbf{x}_i=\varphi(\zeta^{-1}(u_{i,1}),\ldots,\zeta^{-1}(u_{i,d/2}),u_{i,d/2+1},\ldots,u_{i,d})$ where 
            $$\zeta:\mathcal{R}\mapsto\mathcal{R},\quad \zeta(s)=\int_0^s\sqrt{1+\pi^2 u^2}du,$$
            and $(u_{i,1},\ldots,u_{i,d})^\dT\sim \text{MVN}(\boldsymbol{\mu}_{\textbf{x}},\boldsymbol{\Sigma}_{\textbf{x}})$. For $i=1,\ldots,n_0$, we generate $\textbf{y}_i$ in the same way except that $(u_{i,1},\ldots,u_{i,d})^\dT\sim \text{MVN}(\boldsymbol{\mu}_{\textbf{y}},\boldsymbol{\Sigma}_{\textbf{y}}).$ Values of $\boldsymbol{\mu}_{\textbf{x}},\boldsymbol{\Sigma}_{\textbf{x}}$ and $\boldsymbol{\mu}_{\textbf{y}},\boldsymbol{\Sigma}_{\textbf{y}}$ under different scenarios are summarized as follows.
\begin{itemize}
			\item Null: $\boldsymbol{\mu}_{\textbf{x}}=\boldsymbol{\mu}_{\textbf{y}}=\mathbf{0}_d$ and $\boldsymbol{\Sigma}_{\textbf{x}}=\boldsymbol{\Sigma}_{\textbf{y}}=\textbf{I}_{d\times d}$.
			\item Location alternative: $\boldsymbol{\mu}_{\textbf{x}}=0.06\times \mathbf{1}_d$, $\boldsymbol{\mu}_{\textbf{y}}=-0.06\times \mathbf{1}_d$ and $\boldsymbol{\Sigma}_{\textbf{x}}=\boldsymbol{\Sigma}_{\textbf{y}}=\textbf{I}_{d\times d}$.
			\item Direction alternative: $\boldsymbol{\mu}_{\textbf{x}}=\boldsymbol{\mu}_{\textbf{y}}=\mathbf{0}_d$, $$\boldsymbol{\Sigma}_{\textbf{x}}=\textbf{I}_5\otimes\begin{pmatrix}
			    1&0.12\\
                0.12&1
			\end{pmatrix}\text{ and }\boldsymbol{\Sigma}_{\textbf{y}}=\textbf{I}_{d/2}\otimes\begin{pmatrix}
			    1&-0.12\\
                -0.12&1
			\end{pmatrix},$$
            where $\otimes$ stands for the Kronecker product.
		\end{itemize}

Table \ref{Manifold} presents rejection proportions of all tests under different scenarios. Most of the estimated $\hat{d}$ are slightly overestimated as $11$ or $12$. Under the null, the rejection proportions of the Delaunay weighted test are close to the target significance levels, suggesting its ability to control the type I error rate. Although the Delaunay weighted test is not the most powerful under the location alternative, its power is comparable to most of the existing tests except the kernel test, which performs the best under the location alternative. Under the direction alternative, the Delaunay weighted test has the dominating power. Such an advantage even amplifies when sample sizes grow. The covariance test, designed to detect the directional difference, performs the second best under the direction alternative.  In summary, when the data lie on a manifold, our Delaunay weighted test is able to detect general distribution differences with emphasis on the difference in the principal directions.

		\renewcommand{\arraystretch}{0.9}
		\begin{table}[t]
			\centering
			\caption{Rejection proportions (\%) of different two-sample tests on the manifold data under the null and different alternatives ($H_1$) and different significance levels $\alpha=0.01,0.05,0.10$, where the best performance under $H_1$ is highlighted in boldface.}
			\label{Manifold}
			\resizebox{\columnwidth}{!}{
				\begin{tabular}{lrrrrrrrrrrr}
					\toprule
					\multirow{2}*{Tests}&\multicolumn{3}{c}{Null ($\alpha$)}&&\multicolumn{3}{c}{Location $H_1$ ($\alpha$)}&&\multicolumn{3}{c}{Direction $H_1$ ($\alpha$)}\\
					\cline{2-4}\cline{6-8}\cline{10-12}
					&{$0.01$}&{$0.05$}&{$0.10$}&&{$0.01$}&{$0.05$}&{$0.10$}&&{$0.01$}&{$0.05$}&{$0.10$}\\
					\midrule
					&\multicolumn{11}{c}{$n_1=250$, $n_0=250$}\\
					\cline{2-12}
					Delaunay weighted&0.6&5.4&10.7&&9.1&23.6&34.7&&\textbf{14.4}&\textbf{34.1}&\textbf{46.6}\\
Kernel&0.7&4.9&9.7&&\textbf{19.8}&\textbf{42.9}&\textbf{57.8}&&3.5&15.9&28.5\\
Regression&0.4&4.1&9.3&&9.9&27.5&36.0&&2.3&11.8&21.8\\
$k$-NN&0.9&5.6&9.4&&9.9&24.3&35.4&&3.7&12.0&20.9\\
$k$-MST&1.7&5.7&11.6&&9.2&18.1&27.9&&3.5&12.1&18.2\\
$e$-distance&0.9&4.9&10.4&&14.8&26.3&40.5&&1.4&6.8&12.8\\
Covariance&1.1&3.6&6.7&&0.2&1.5&4.4&&11.5&31.0&42.5\\
					
                    &\multicolumn{11}{c}{$n_1=500$, $n_0=500$}\\
					\cline{2-12}
					Delaunay weighted&0.6&5.7&10.8&&29.2&44.0&58.1&&\textbf{46.7}&\textbf{69.7}&\textbf{81.4}\\
Kernel&1.9&5.2&9.2&&\textbf{62.6}&\textbf{80.3}&\textbf{86.8}&&11.1&35.0&50.5\\
Regression&1.9&5.9&11.1&&32.7&54.1&66.8&&7.7&21.0&34.7\\
$k$-NN&1.9&5.7&10.0&&27.5&47.2&60.2&&5.7&22.1&34.7\\
$k$-MST&1.3&6.4&10.2&&20.1&38.4&49.2&&12.0&27.8&36.3\\
$e$-distance&1.6&6.0&10.2&&30.2&47.8&59.0&&1.7&7.5&15.5\\
Covariance&1.3&3.5&5.9&&1.0&3.7&6.2&&43.4&66.9&78.2\\
					\bottomrule
				\end{tabular}
			}
		\end{table} 
		
\subsubsection{Sensitivity Analysis}\label{Sensitivity}

As described in Section \ref{isomap}, a manifold learning procedure is required in our Delaunay weighted test on manifold data, including estimating the intrinsic dimension, estimating the geodesic distance, and obtaining the low-dimensional Euclidean representation $\widetilde{\mathbb{Z}}$ via the MDS. It is of interest to see how the manifold learning steps affect the performance of our Delaunay weighted test. Specifically, we generate $1000$ datasets under both the null and the direction alternative as in Section \ref{Size_Power} with sample size $n_1=n_0=500$. We consider three implementations of our test for comparison.
\begin{itemize}
    \item Known Manifold: We implement the Delaunay weighted test based on the known manifold structure. That is to say, we directly calculate the Delaunay weights using Definition \ref{DW} and perform the permutation test.
    \item Fixed $\hat{d}$: We implement the Delaunay weighted test by fixing $\hat{d}$ at several values including the true intrinsic dimension. 
    \item Estimated $\hat{d}$: We implement the Delaunay weighted test with all the manifold learning steps. 
\end{itemize}

\begin{figure}[t]
			\centering
			\begin{subfigure}{0.32\linewidth}
				\includegraphics[width=\linewidth]{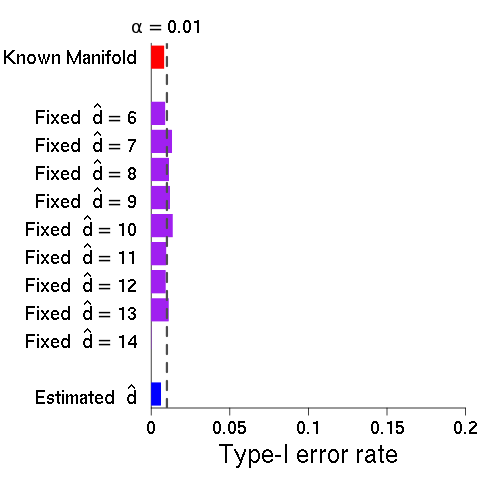}
				\caption{Null with $\alpha=0.01$.}
			\end{subfigure}
			\begin{subfigure}{0.32\linewidth}
				\includegraphics[width=\linewidth]{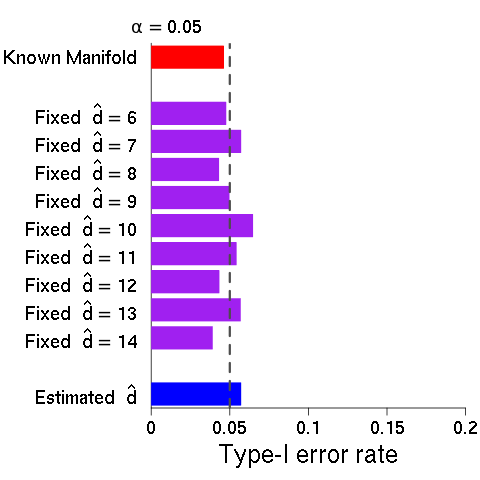}
				\caption{Null with $\alpha=0.05$.}
			\end{subfigure}
			\begin{subfigure}{0.32\linewidth}
				\includegraphics[width=\linewidth]{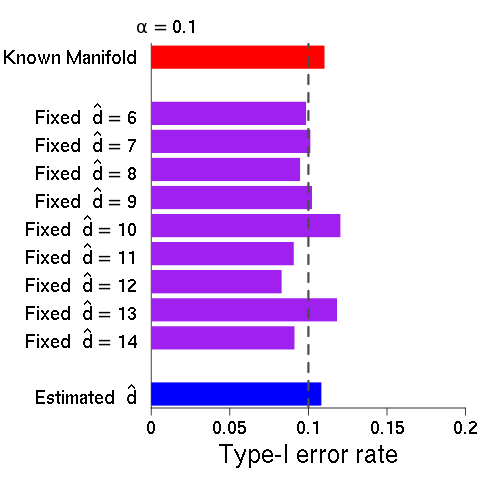}
				\caption{Null with $\alpha=0.10$.}
			\end{subfigure}
            \begin{subfigure}{0.32\linewidth}
				\includegraphics[width=\linewidth]{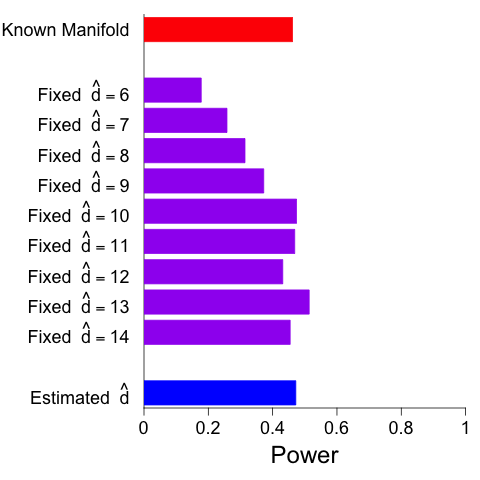}
				\caption{Direction $H_1$ with $\alpha=0.01$.}
			\end{subfigure}
			\begin{subfigure}{0.32\linewidth}
				\includegraphics[width=\linewidth]{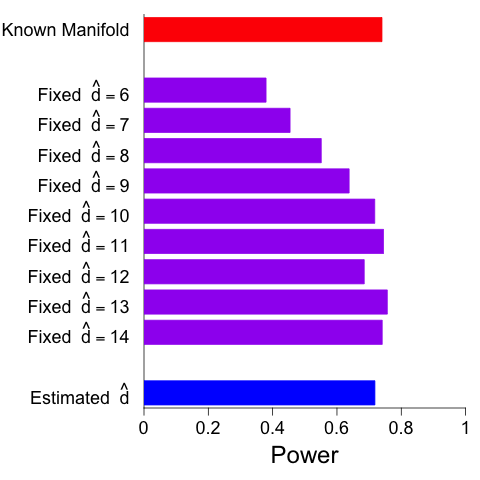}
				\caption{Direction $H_1$ with $\alpha=0.05$.}
			\end{subfigure}
			\begin{subfigure}{0.32\linewidth}
				\includegraphics[width=\linewidth]{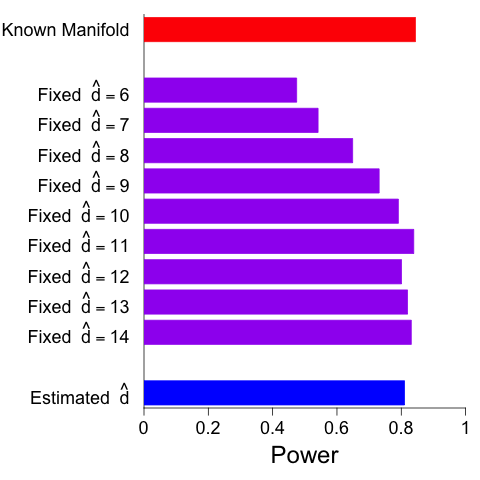}
				\caption{Direction $H_1$ with $\alpha=0.10$.}
			\end{subfigure} 
			\caption{Rejection proportions of different implementations of the Delaunay weighted test under the null and the direction alternative as in Section \ref{Size_Power} with sample size $n_1=n_0=500$ and different significance levels ($\alpha$).}
			\label{fig:Sensitivity} 
		\end{figure}

Figure \ref{fig:Sensitivity} displays the rejection proportions of different implementations of the Delaunay weighted test under different scenarios. In general, the Delaunay weighted test manages to control the type I error rate under the null and the power does not vary much with known or unknown manifold structures, indicating the Delaunay weighted test is robust to the manifold learning steps. This is corroborated by the similar results among the cases of ``Known Manifold", ``Fixed $\hat{d}=10=d$", and ``Estimated $\hat{d}$". Because our goal is to test the hypothesis rather than accurately estimate the manifold structure, the Delaunay weighted test only loses power if the estimated intrinsic dimension $\hat{d}$ is lower than the truth. Indeed, as shown in the results of  ``Fixed $\hat{d}$" with different values of $\hat{d}$, we see that the power keeps increasing as $\hat{d}$ grows from 6 to 10 and does not fluctuate drastically for a small range of $\hat{d}> 10$. Recall from Section \ref{Size_Power} that the intrinsic dimension $d$ tends to be slightly overestimated for this example. It is as expected that the implementation ``Estimated $\hat{d}$'' shows robust performance. 

\subsubsection{Computational Cost}\label{Computation}		
		To investigate the computational cost of the Delaunay weighted test compared with existing approaches, we generate $200$ datasets under the null in a way similar  to Section \ref{Size_Power} with different sample sizes $n=n_1+n_0$ ($n_1=n_0$), intrinsic dimensions $d$ and ambient dimensions $D$. Table \ref{Time} summarizes the computational time of different steps in the Delaunay weighted test and existing approaches. In general, the computational time of the Delaunay weighted test is mainly dominated by the computational time in estimating the geodesic distance and calculating the Delaunay weight matrix. More specifically, the computational time in calculating the Delaunay weight matrix grows as the sample size and the intrinsic dimension increase, while the computational time in estimating the geodesic distance only depends on the sample size.  Although the Delaunay weighted test is only more computationally efficient than the $k$-MST test, which requires an iterative procedure, its computational cost is still affordable.  

		\begin{sidewaystable}[!htbp]
			\centering
			\caption{Mean (standard deviation) of the computational time (unit: s) of different steps in the Delaunay weighted test and existing approaches under different scenarios.}
			\label{Time}
			\resizebox{\columnwidth}{!}{
				\begin{tabular}{lrrrrrrrr}
					\toprule
                    &\multicolumn{3}{c}{$d=10,D=100$}&&\multicolumn{4}{c}{$d=10,n=1000$}\\
                    \cline{2-4}\cline{6-9}
                    &$n=250$&$n=500$&$n=1000$&&$D=50$&$D=100$&$D=150$&$D=200$\\
                    \midrule
					Algorithm S1&0.44(0.24)&1.18(0.29)&4.00(0.92)&&3.51(0.81)&4.00(0.92)&4.36(0.87)&4.59(0.84)\\
Estimating $\tilde{d}_{\mathcal{M}}(\textbf{z}_i,\textbf{z}_j)$&3.52(0.32)&17.92(0.67)&112.20(6.34)&&110.64(5.47)&112.20(6.34)&112.27(6.61)&110.43(5.65)\\
MDS&0.01(0.00)&0.05(0.01)&0.30(0.04)&&0.29(0.04)&0.30(0.04)&0.29(0.04)&0.28(0.01)\\
Calculating $\boldsymbol{\Gamma}_{\widetilde{\mathbb{Z}}}$&36.83(5.40)&140.27(11.66)&552.19(35.54)&&550.21(40.17)&552.19(35.54)&560.09(31.87)&544.86(32.19)\\
Permutation test&0.39(0.04)&0.42(0.02)&0.52(0.02)&&0.52(0.03)&0.52(0.02)&0.52(0.01)&0.51(0.01)\\
Total&41.20(5.99)&159.84(12.66)&669.20(42.86)&&665.18(46.52)&669.20(42.86)&677.54(39.41)&660.67(38.70)\\\midrule
$k$-NN test&15.88(1.99)&17.31(1.12)&22.69(1.17)&&22.17(1.19)&22.69(1.17)&23.45(1.00)&22.99(1.02)\\
$k$-MST test&36.80(4.30)&201.67(13.59)&1327.99(71.17)&&1321.71(74.25)&1327.99(71.17)&1340.22(59.30)&1299.28(66.24)\\
Kernel test&0.40(0.05)&1.38(0.15)&4.79(0.48)&&4.85(0.50)&4.79(0.48)&4.95(0.59)&4.77(0.50)\\
$e$-distance test&0.45(0.05)&1.74(0.16)&5.75(0.70)&&5.52(0.61)&5.75(0.70)&5.84(0.73)&5.80(0.69)\\
Covariance test&0.19(0.28)&0.29(0.43)&0.48(0.31)&&0.20(0.28)&0.48(0.31)&0.82(0.32)&1.20(0.33)\\
Regression test&0.51(0.02)&1.63(0.07)&5.25(0.29)&&4.81(0.23)&5.25(0.29)&5.77(0.36)&6.09(0.38)\\
\midrule
&\multicolumn{6}{c}{$D=100,n=1000$}&&\\
                    \cline{2-7}
                    &$d=2$&$d=4$&$d=6$&&$d=8$&$d=10$&&\\\midrule
Algorithm S1&3.82(0.49)&3.80(0.57)&3.87(0.67)&&3.88(0.76)&4.00(0.92)&&\\
Estimating $\tilde{d}_{\mathcal{M}}(\textbf{z}_i,\textbf{z}_j)$&110.42(3.94)&110.39(4.92)&110.16(5.08)&&110.67(5.59)&112.20(6.34)&&\\
MDS&0.31(0.04)&0.30(0.06)&0.27(0.02)&&0.29(0.04)&0.30(0.04)&&\\
Calculating $\boldsymbol{\Gamma}_{\widetilde{\mathbb{Z}}}$&46.25(1.42)&145.26(15.96)&296.68(17.11)&&442.41(31.65)&552.19(35.54)&&\\
Permutation test&0.47(0.03)&0.50(0.03)&0.51(0.01)&&0.52(0.03)&0.52(0.02)&&\\
Total&161.27(5.91)&260.25(21.54)&411.48(22.89)&&557.77(38.07)&669.2(42.86)&&\\\midrule
$k$-NN test&8.47(0.28)&11.40(0.59)&15.72(0.64)&&19.59(1.05)&22.69(1.17)&&\\
$k$-MST test&322.64(8.94)&544.54(38.21)&855.25(38.38)&&1127.50(62.76)&1327.99(71.17)&&\\
Kernel test&5.10(0.57)&5.03(0.58)&4.69(0.31)&&4.62(0.30)&4.79(0.48)&&\\
$e$-distance test&5.76(1.13)&5.90(1.08)&5.67(0.94)&&5.56(0.85)&5.75(0.70)&&\\
Covariance test&0.44(0.30)&0.51(0.33)&0.62(1.36)&&0.49(0.34)&0.48(0.31)&&\\
Regression test&15.15(1.27)&7.14(0.48)&5.92(0.32)&&5.40(0.30)&5.25(0.29)&&\\
					\bottomrule
                    \multicolumn{9}{l}{{\footnotesize CPU: AMD EPYC 9754, 2.25GHz.}}
				\end{tabular}
			}
		\end{sidewaystable} 
		
\subsection{Real Data Analysis}\label{Human_Face}
        
As an illustration, we apply the Delaunay weighted test to the 
mice protein expression dataset\footnote{\url{https://archive.ics.uci.edu/dataset/342/mice+protein+expression}} \citep{Higuera2015}. This dataset records the expression levels of 77 proteins in 1080 cells from the cortex nuclear fraction of 72 mice. These 72 mice belong to 8 groups according to their genotype (control or trisomic), behavior (stimulated to learn or not) and treatment (injected with the drug ``memantine" or only saline). We remove 3 cells and 9 proteins with more than 10 missing values, resulting in a complete dataset of 1077 cells and $D=68$ proteins.

\begin{figure}[t]
			\centering
			\begin{subfigure}{0.32\linewidth}
				\includegraphics[width=\linewidth]{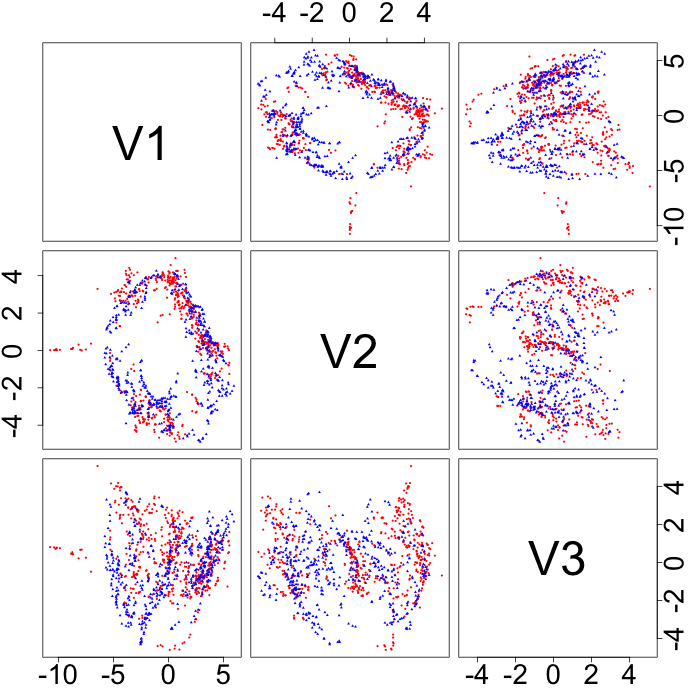}
                \includegraphics[width=\linewidth]{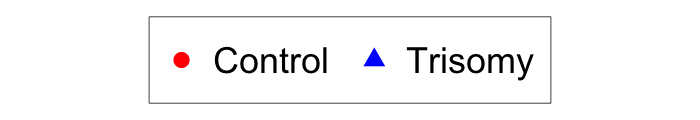}
				\caption{Genotype}
			\end{subfigure}
            \begin{subfigure}{0.32\linewidth}
				\includegraphics[width=\linewidth]{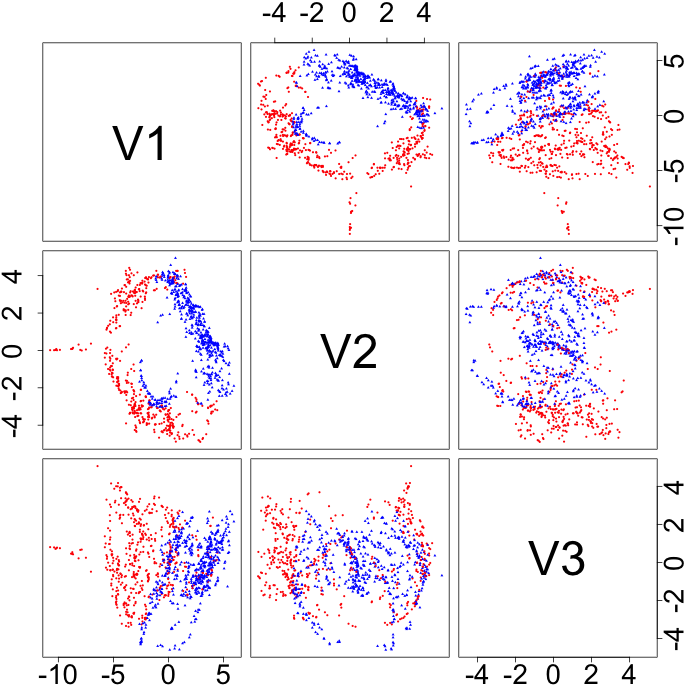}
                \includegraphics[width=\linewidth]{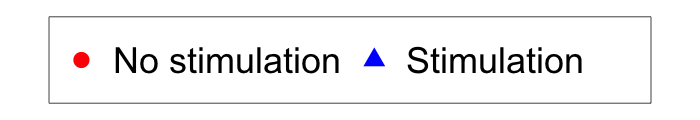}
				\caption{Behavior}
			\end{subfigure}
			\begin{subfigure}{0.32\linewidth}
				\includegraphics[width=\linewidth]{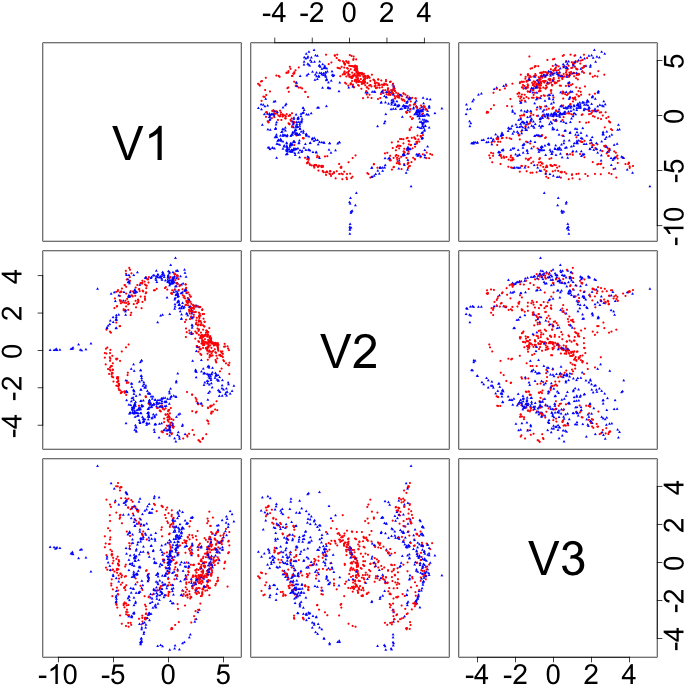}
                \includegraphics[width=\linewidth]{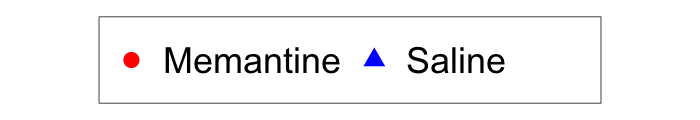}
				\caption{Treatment}
			\end{subfigure} 
			\caption{Low-dimensional Euclidean representations of protein expression levels of 1077 cells from the mouse cortex, where V$k$ denotes the $k$-th coordinate of the Euclidean representations, and cells are colored according to their (a) genotypes, (b) behaviors and (c) treatments.}
			\label{fig:Mice}
		\end{figure}

To check whether or not the protein expression levels are likely supported on a low-dimensional manifold, we apply Algorithm S1 \citep{Facco2017} to estimate the intrinsic dimension using all cells. This gives the estimated intrinsic dimension $\hat{d}=3$, which is much smaller than $D=68$. We further compute the low-dimensional Euclidean representations of all cells via the MDS, as visualized in Figure \ref{fig:Mice}. The protein expression level difference is prominent between cells of different behaviors, while the difference is less obvious with respect to different genotypes and treatments. Thus, we focus on the two hypothesis tests:
\begin{itemize}
    \item $H_{0,\text{g}}:$ No difference in protein expression levels between cells of different genotypes.
    \item $H_{0,\text{t}}:$ No difference in protein expression levels between cells of different treatments.
\end{itemize}

It is likely that protein expression levels are correlated for cells of the same mouse. Thus, a direct permutation test
on all the cells may not be consistent.  Instead, we perform the Delaunay weighted test with the mouse-level permutation, under the assumption that cells of different mice are independent. Due to possible correlation, bias may be introduced in computing the low-dimensional representations as well as in the Delaunay weights, which would incur power loss. However, the type I error is controlled by using the mouse-level permutation. Similarly, all existing methods under the comparison are also performed with the mouse-level permutation.

We compute $p$-values of our Delaunay weighted test and all other approaches in testing $H_{0,\text{g}}$ and $H_{0,\text{t}}$, as shown  in Table \ref{Real_data}. We find that all the tests except the $e$-distance test are significant with respect to $H_{0,\text{g}}$ under $\alpha=0.05$. This suggests different protein expression levels between mice of different genotypes, as expected. However, in testing $H_{0,\text{t}}$, only our Delaunay weighted test yields a significant result under $\alpha=0.05$, while 
all other tests fail to reject the null.
This indicates that the drug ``memantine" may cause changes in protein expression of mouse cortex cells, which can be detected by the tests that exploit geometric structures. 

\renewcommand{\arraystretch}{0.9}
    \begin{table}[t]
        \centering
        \caption{ The $p$-values of different two-sample tests in testing $H_{0,\text{g}}$ and $H_{0,\text{t}}$. Here, ``$^{**}$" denotes significance under $\alpha=0.05$ and ``$^{*}$" denotes significance under $\alpha=0.10$.}
        \label{Real_data}
        \begin{tabular}{lll}
             \toprule
Tests&$H_{0,\text{g}}$&$H_{0,\text{t}}$\\
             \midrule
Delaunay weighted&0.028$^{**}$&0.011$^{**}$\\
Kernel&0.012$^{**}$&0.076$^{*}$\\
Regression&0.010$^{**}$&0.776\\
$k$-NN&0.038$^{**}$&0.501\\
$k$-MST&0.021$^{**}$&0.880\\
$e$-distance&0.992&0.806\\
\bottomrule
        \end{tabular}
    \end{table} 
    
        Based on the discovery by the Delaunay weighted and kernel tests that the drug ``memantine" induces different protein expression, we further investigate which subgroups of mice the drug is effective on. Specifically, we perform a stratified analysis of $H_{0,\text{t}}$ over mice subgroups defined by genotypes and behaviors and report the results in Table \ref{Real_data2}. We find that both the Delaunay weighted and kernel tests detect expression level changes of trisomic mice who are not stimulated. However, the Delaunay weighted test additionally detects a significant difference (under $\alpha=0.05$) for the trisomic and stimulated mice. This is the main reason why the Delaunay weighted test has a smaller $p$-value than the kernel test does with respect to $H_{0,\text{t}}$ in Table \ref{Real_data}.

        \renewcommand{\arraystretch}{0.9}
    \begin{table}[t]
        \centering
        \caption{  The $p$-values of the Delaunay weighted test and the kernel test in testing $H_{0,\text{t}}$ over different subgroups. Here, ``$^{**}$" denotes significance under $\alpha=0.05$.}
        \label{Real_data2}
        \begin{tabular}{lllll}
             \toprule
             \multirow{2}*{Tests}&Control \&&Control \&&Trisomic \&&Trisomic \&\\
&stimulated&not stimulated&stimulated&not stimulated\\
             \midrule
Delaunay weighted&0.126&0.245&0.037$^{**}$&0.005$^{**}$\\
Kernel&0.252&0.171&0.135&0.002$^{**}$\\
\bottomrule
        \end{tabular}
    \end{table}

		\section{Discussions}\label{Discussions}
		
		We propose a new two-sample test for high-dimensional data by incorporating geometric information on manifolds. Based on the Delaunay triangulation on manifolds, a new geometric proximity measure named the Delaunay weight is developed with a series of computational strategies for efficient calculation. The Delaunay weight distinguishes itself from others by taking the direction information into account, and thus the resulting test is more efficient in detecting the direction difference of two distributions. We define the test statistic as the average of within-group Delaunay weights, whose $p$-value is computed via the permutation procedure. Theoretical derivation shows that under mild conditions, the test statistic is asymptotically normal under the null and the proposed test is consistent against general alternatives. Experiments demonstrate broad applicability and advantages of the Delaunay weighted test compared with existing ones when two distributions differ in their principal directions of covariance matrices.

        There are several directions for future work. First, as the type of alternative is usually unknown in real applications and our Delaunay weighted test is particularly powerful in detecting directional differences, it is inspiring to combine our test with the kernel or $e$-distance test through a Cauchy combination test \citep{Liu2019} for more robust performance across different types of distributional differences. Second, it is of interest to utilize the Delaunay weight matrix $\boldsymbol{\Gamma}_{\mathbb{Z}}$ that captures pairwise geometric information for more complicated statistical inferences other than hypothesis testing. Finally, it is interesting to generalize the Delaunay weight framework to the case of noise-contaminated data or data residing on a union of manifolds of different dimensions. Such extensions would substantially enhance the scope of our methodology, enabling its application to more complex real-world datasets.

\section*{Code Availability}

The R code for reproducing all the experiments is available at \url{https://github.com/GuJQ5/Delaunay-weighted-test}.

\section*{Acknowledgments}
We thank the Editor, the Associate Editor and two referees for their careful reviews and many insightful comments, which led to a much better exposition of our work. Tan's research was supported by the National Natural Science Foundation of China (No.~12401363 and 12471263), the Fundamental Research Funds for the Central Universities, and the Key Laboratory of Intelligent Computing and Applications (Ministry of Education). Yin's research was  supported by the Patrick SC Poon endowment fund.

\spacingset{1}
\bibliographystyle{apalike}
\bibliography{thesis_new}

\end{document}